\documentclass[11pt,sort&compress]{elsarticle}

\makeatletter
  \long\def\pprintMaketitle{\clearpage
  \iflongmktitle\if@twocolumn\let\columnwidth=\textwidth\fi\fi
  \resetTitleCounters
  \def\baselinestretch{1}%
  \printFirstPageNotes
  \begin{center}%
 \thispagestyle{pprintTitle}%
   \def\baselinestretch{1}%
    {\large\bf\@title}\par\vskip5pt
    \normalsize\elsauthors\par\vskip5pt
    \footnotesize\itshape\elsaddress\par\vskip10pt
    \end{center}%
  \gdef\thefootnote{\arabic{footnote}}%
  }
\makeatother

\usepackage{lmodern}
\usepackage[utf8]{inputenc}
\usepackage[T1]{fontenc}
\usepackage{amsmath}
\usepackage{amsfonts}
\usepackage{lipsum}
\usepackage{graphicx}
\usepackage{xspace}
\usepackage{epsf}
\usepackage{setspace}
\usepackage{float}
\usepackage{soul}
\usepackage[
    protrusion=true,
    activate={true,nocompatibility},
    final,
    tracking=true,
    kerning=true,
    spacing=true,
    factor=1100]{microtype}
\SetTracking{encoding={*}, shape=sc}{40}

\usepackage[margin=1in]{geometry}
\usepackage{abstract}
\usepackage{booktabs}

\usepackage{hyperref}
\usepackage{graphicx}

\usepackage{stmaryrd}
\SetSymbolFont{stmry}{bold}{U}{stmry}{m}{n}

\usepackage{placeins}
\usepackage{mathrsfs}

 \usepackage{bm}
  \usepackage{subfig}
  \usepackage{caption}
\usepackage{algorithm}
\usepackage{algpseudocode}
\usepackage{url}
\usepackage[parfill]{parskip}

   \usepackage{graphicx}
 \usepackage{float}
\usepackage{epstopdf, epsfig}
\usepackage{physics}

\usepackage[]{xcolor}
\usepackage{stmaryrd}
\usepackage{graphics}
\usepackage{appendix}

\usepackage{kantlipsum}

\makeatletter
\renewcommand{\Function}[2]{%
  \csname ALG@cmd@\ALG@L @Function\endcsname{#1}{#2}%
  \def\jayden@currentfunction{#1}%
}
\newcommand{\funclabel}[1]{%
  \@bsphack
  \protected@write\@auxout{}{%
    \string\newlabel{#1}{{\jayden@currentfunction}{\thepage}}%
  }%
  \@esphack
}

 \usepackage{amsmath,amssymb}

\begin{document}

\hypersetup{
  pdfauthor=author,
}

\begin{frontmatter}
\title{Stochastic reduced-order Koopman model for turbulent flows}


\author[add1]{Tianyi Chu\corref{cor1}}
\ead{tic173@ucsd.edu}
\author[add1]{Oliver T. Schmidt}

\address[add1]{Department of Mechanical and Aerospace Engineering,  University of California San Diego, 9500 Gilman Drive, La Jolla, CA 92093, USA}
\cortext[cor1]{Corresponding author}
    \date{}

\end{frontmatter}

\begin{abstract}

A stochastic data-driven reduced-order model applicable to a wide range of turbulent natural and engineering flows is presented. Combining ideas from Koopman theory and spectral model order reduction, the stochastic low-dimensional inflated convolutional Koopman model (SLICK) accurately forecasts short-time transient dynamics while preserving long-term statistical properties. A discrete Koopman operator is used to evolve convolutional coordinates that govern the temporal dynamics of spectral orthogonal modes, which in turn represent the energetically most salient large-scale coherent flow structures. Turbulence closure is achieved in two steps: first, by inflating the convolutional coordinates to incorporate nonlinear interactions between different scales, and second, by modeling the residual error as a stochastic source. An empirical dewhitening filter informed by the data is used to maintain the second-order flow statistics within the long-time limit. The model uncertainty is quantified through either Monte Carlo simulation or by directly propagating the model covariance matrix. The model is demonstrated on the Ginzburg-Landau equations, large-eddy simulation (LES) data of a turbulent jet, and particle image velocimetry (PIV) data of the flow over an open cavity. In all cases, the model is predictive over time horizons indicated by a detailed error analysis and integrates stably over arbitrary time horizons, generating realistic surrogate data. 

\end{abstract}


\section{Introduction}\label{Sec:intro}

The complex and chaotic dynamics, coupled with a wide range of length and time scales, remain an open challenge for the reduced-order modeling of turbulent flows.
Despite the understanding of the Navier-Stokes equations and the availability of accurate high-fidelity 
 methods to solve them, real-time prediction of turbulent flows remains elusive across various fields including aero-
 and hydrodynamics, physical oceanography, atmospheric data, and technical engineering flows, primarily due to their immense computational costs.
Available applications, therefore, have often been demonstrated using simplified dynamical systems, such as the Ginzburg-Landau equation and the Lorenz system, or canonical problems like cylinder wakes. This work aims to develop a reduced-order model (ROM) applicable to real-world data.

ROMs have become a prevalent approach in fluid mechanics to 
reduce complexity by separating temporal dynamics that are 
governed
 by simplified equations, such as ordinary differentiation equations (ODEs), from the spatial flow features. 
The latter spatial features are compressed into a modal basis of spatial fields, denoted as $\{\vb*{\psi}_j\}_{j=1}^{N_k}$.
Practical applications of ROMs for turbulent flows include real-time simulations \citep{carlberg2013gnat, kaiser2014cluster, brunton2020machine}, flow control \citep{ kim2007linear, sipp2010dynamics,rowley2017model,zare2017colour}, and uncertainty quantification (UQ) \citep{sapsis2013blended, sapsis2013statistically,majda2018model}.
Model reduction significantly decreases the computational complexity and enables a better understanding of the underlying dynamics.
Arguably, the most successful ROMs in fluid mechanics are based on Galerkin projection methods, which project the high-dimensional governing equations onto the subspace of spatial modes to get coupled ODEs.
Readers are referred to Benner \textit{et al.} \cite{benner2015survey} for a review of projection-based methods.
These operator-based ROMs require knowledge of the governing equations. Therefore, they are challenged when the physical description of the system dynamics is difficult to formulate, as is true in examples including climate science, physiology, and fluid systems with only partial measurements. As an alternative, data-driven modeling techniques have become increasingly popular for analyzing and controlling high-dimensional dynamical systems due to the rise of high-fidelity data collection, see, e.g., \cite{brunton2015closed,kutz2016dynamic,duriez2017machine}.




\begin{figure*}[ht!]
		\centering
        \includegraphics[trim = 10mm 0mm 5mm 5mm,width=1\textwidth]{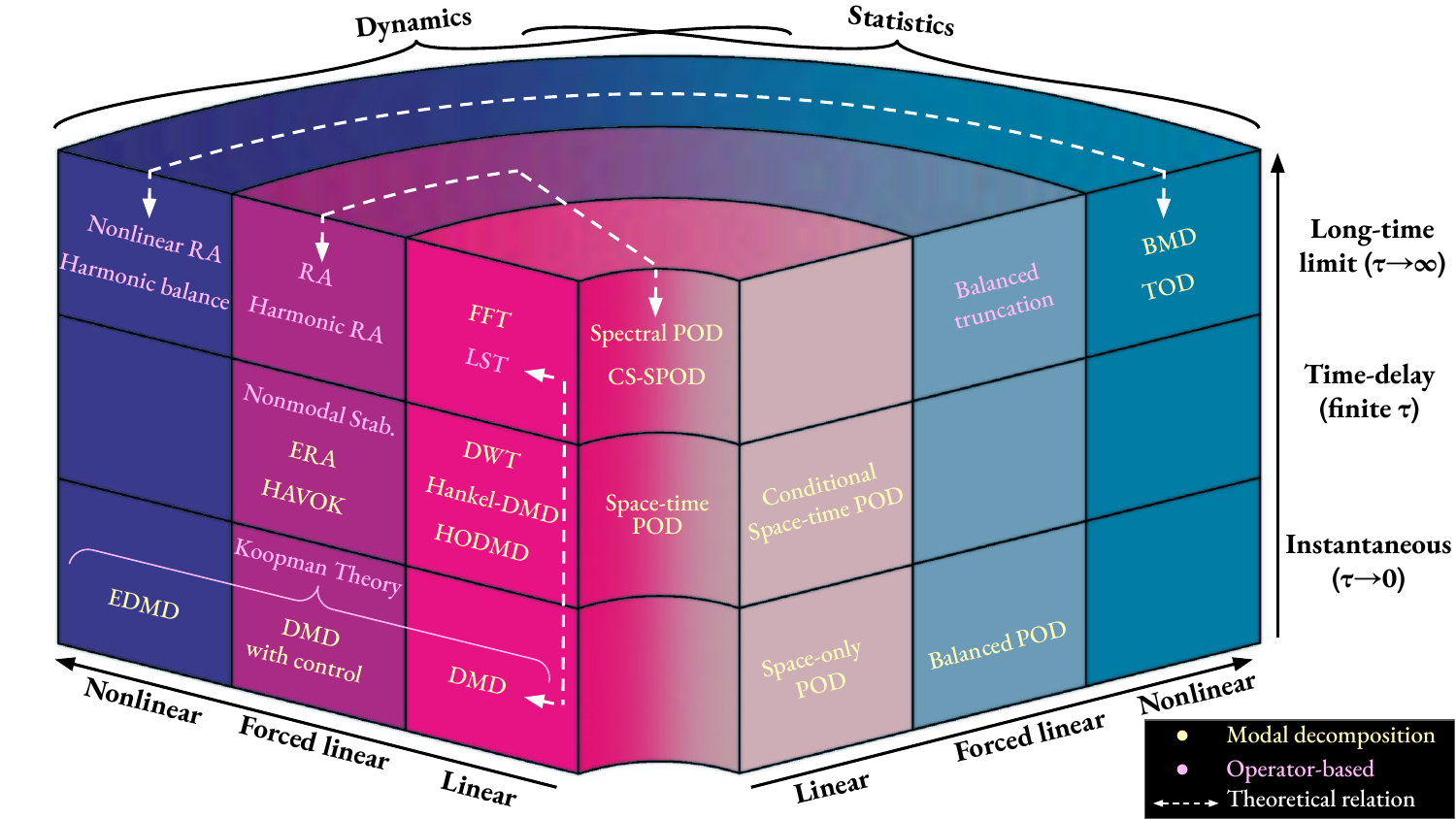} 
        \caption{Classification of 
        data-driven and operator-based methods for basis identification methods for model order reduction, categorized according to three modeling perspectives: 
        (i) analysis type (statistical, dynamical, or hybrid), (ii) state vector structure (linear $\to$ forced linear $\to$ nonlinear), and (iii) time-delay horizon (instantaneous $\to$ finite-time $\to$ long-time limit).
        }  \label{Modal_schematic}
	\end{figure*}

 
The choice of model order reduction usually goes hand-in-hand with the modal basis. 
Figure~\ref{Modal_schematic} categorizes established data-driven and operator-based theoretical basis identification methods for model order reduction from three modeling perspectives.
Note that all the methods here are applied to analyze general nonlinear fluid flows.
The terminology 'linear,' 'forced linear,' and 'nonlinear' pertains to the structure of the state vector under analysis. 
Detailed interpretations are provided in the Supplementary Material.
In fluid mechanics, the most widely used approach for extracting the dominant features from a time-series data is proper orthogonal decomposition (POD); see Lumley \cite{lumley1967structure,lumley2007stochastic}. 
In other disciplines, POD is known as empirical orthogonal functions (EOF), principal component analysis (PCA), or Karhunen-Lo\`eve decomposition.
POD modes are particularly suitable for Galerkin projection due to their 
optimality and orthogonality; see, e.g., \cite{noack2003hierarchy,rowley2004model}.
Most commonly, the classical POD modes are computed using the method of snapshots by Sirovich \cite{sirovich1987turbulence} through the eigendecomposition of the spatial cross-correlation tensor.

As an alternative to POD, dynamic mode decomposition (DMD) \citep{schmid2010dynamic} aims at identifying coherent spatial modes that best describe the flow dynamics upon linearization.
The temporal dynamics of the coefficients are described by complex exponentials, specifically involving frequency and growth/decay rate.
The DMD method provides a spatiotemporal decomposition of data streams and is closely related to Koopman analysis \citep{mezic2005spectral, rowley2009spectral,mezic2013analysis}.
The underlying idea of DMD is to approximate the infinite-dimensional linear operator that governs the evolution of observables of a nonlinear system, known as the Koopman operator \citep{koopman1931hamiltonian}, on a finite manifold.
The most popular flavor of DMD is the so-called exact DMD method \citep{Tu2014}, which is based on the eigendecomposition of the finite-dimensional DMD operator.
The DMD modes, also called dynamic modes, reduce to temporal discrete Fourier
transform (DFT) modes if the data is periodic \citep{rowley2009spectral} and approximate the Koopman eigenfunctions if snapshots are independent \citep{chen2012variants}.
Formally, Koopman theory requires the definition of observables of the flow state. 
In practice, the full flow state is often chosen as observables for DMD.
As a generalization of the standard DMD, Williams \textit{et al.} \cite{williams2015data} proposed the extended DMD (EDMD) method for approximating the leading Koopman eigenvalues and eigenfunctions by using a dictionary of scalar observables.
In EDMD, every function of the flow state can be considered as observables, in particular, nonlinear observable functions, such as kernel functions \citep{Williams2014}, Hermite polynomials \citep{li2017extended}, and candidate functions obtained using sparse regression \citep{brunton2016koopman}.
Koopman theory guarantees the convergence of the EDMD operator to the Koopman operator in the limit of infinite observables \citep{korda2018convergence}. While (E)DMD focuses on embedding the state dynamics onto (non)linear manifolds in space, the time-delay embedding of data sequences provides a different perspective for identifying dominant spatiotemporal structures. Arbabi $\&$ Mezi\'c \cite{arbabi2017ergodic} introduced the Hankel-DMD algorithm to obtain the Koopman spectrum by performing the DMD of a Hankel data matrix. 
This Hankelized approach is guaranteed to retrieve Koopman eigenfunctions for ergodic systems when an infinite amount of data is available. Le Clainche $\&$ Vega \cite{le2017hodmd} proposed the higher-order DMD (HODMD) to address high spectral complexity.
This approach has been successfully demonstrated on a range of flows characterized by a broadband spectrum, including a jet \citep{le2017higher_expdata}, a cylinder wake \citep{le2017higher_3dcylinder}, and an airfoil wake \citep{kou2018reduced}.
The reader is referred to Vega $\&$ Le Clainche \cite{vega2020higher} for more applications.
The above variants of DMD are summarized in the recent review by Schmid \cite{schmid2021dynamic}. In addition to these DMD-based methods, an alternative approach to identifying spatiotemporal structures from data is space-time POD, which originates from the most general version of POD and describes the time evolution of the flow over a specified time window \citep{lumley2007stochastic}.
Schmidt $\&$ Schmid \cite{schmidt2019conditional} proposed the conditional space-time POD to identify acoustic intermittency in a jet.
Recently, Frame $\&$ Towne \cite{frame2023space} demonstrated that discrete space-time POD modes can be obtained through the singular value decomposition (SVD) of a block Hankel matrix.
These Hankel singular vectors form the modal basis of the Hankel alternative view of Koopman (HAVOK) model \citep{brunton2017chaos}. As shown by Frame $\&$ Towne \cite{frame2023space}, they
converge to POD modes in the short-time limit and to spectral POD (SPOD) modes in the long-time limit.


SPOD, the latter variant of POD, identifies large coherent structures in stationary flows.
Analogous to POD, SPOD is computed as the eigendecomposition of the cross-spectral density tensor and decomposes the flow into 
energy-ranked structures that evolve coherently in both space and time
\citep{schmidt2018spectral,towne2018spectral}. 
For the same method, Nekkanti $\&$ Schmidt \cite{nekkanti2021frequency} proposed a convolution-based approach to obtain discrete time-continuous expansion coefficients, $\vb*{a}$, and facilitate time-local analyses, such as frequency-time diagrams.
Besides optimally accounting for the second-order statistics, SPOD modes are also dynamically significant as they are optimally averaged ensemble DMD modes \citep{towne2018spectral} and are formally equivalent to the spectral expansion of the stochastic Koopman operator \citep{mezic2005spectral,arbabi2022generative} for stationary flows.


In this work, we explore and demonstrate the potential of leveraging the aforementioned properties of SPOD to model turbulent flows as a linear time-invariant (LTI) system,

\begin{align}\label{koopman_spod}
   \dv{}{t}\vb*{a} =  \vb*{K} \vb*{a} +\vb*{b},
\end{align}
whose dynamics are inferred in a purely data-driven manner using the Koopman operator theory.
Here, $\vb*{K}$ is a finite-dimensional approximation of the continuous-in-time Koopman operator.
For general fluid systems, using a finite-dimensional linear operator to represent the complex nonlinear dynamics is too restrictive \citep{brunton2016koopman}.
Brunton \textit{et al.} \cite{brunton2016discovering} proposed the sparse identification of nonlinear dynamics (SINDy) method to address the nonlinearity by including additional relevant nonlinear functions.
We here consider an input forcing term, $\vb*{b}$, to account for the unresolved nonlinear interactions and stochastic fluctuations. 
This idea is inspired by the operator-based flow models, such as 
the stochastic structural stability theory (S3T) system \citep{farrell2003structural,farrell2012dynamics},
and the mean-flow based resolvent analysis \cite{hwang2010amplification,mckeon2010critical}. Starting from the control theoretical perspective or system identification,
similar LTI formulations have also been carried out for turbulent systems,
see, e.g., \cite{majda2012physics,sapsis2013statistically, majda2014conceptual,brunton2017chaos}.
The SPOD expansion coefficients, $\vb*{a}$, can be alternatively viewed as time-delay observables of the flow states, specifically through the Fourier convolution. 
This perspective can be understood from the inherent relationship between the Hankel singular vectors and SPOD modes \citep{frame2023space}.
The space-time orthogonality and frequency-mode correspondence of SPOD are retained in the convolution process, and the resulting $\vb*{a}$ is continuously discrete in time. 
Combining Koopman theory with time-delay embedding \cite{takens1981detecting} has been successfully used to capture the nonlinear dynamics of chaotic systems \citep{mezic2004comparison}.
This concept aligns with the principles of
the singular spectral analysis (SSA) \cite{vautard1989singular} and the eigensystem realization algorithm (ERA) \cite{juang1985eigensystem}.
 Applications of the Hankel-DMD method for predicting simple flows 
 include \cite{arbabi2017ergodic,yuan2021flow,eivazi2021recurrent}.
 The closely related HAVOK model \cite{brunton2017chaos} performs linear regression on the principle coordinates of the Hankel matrix.
Its accuracy and the effect of delay horizons have been studied in \cite{kamb2020time,pan2020structure}, showing the effectiveness of linear time-delay models for representing nonlinear dynamics. HODMD similarly provides a linear dynamical system
that governs the evolution of time-lagged states and has been used for
reduced-order modeling, see \citep{kou2018reduced,beltran2022adaptive,le2022data} for recent applications. The common idea behind these linear time-delay models is the SVD of a Hankel matrix. Naturally, their computational cost depends on the spatial and spectral complexities of the data sequence.
In this work, we circumvent this bottleneck by using convolutional coordinates that readily encapsulate time-delay embedding. 
We hence leverage the advantageous properties of convolutional coordinates, theoretically confirmed by Kamb $\textit{et al.}$ \cite{kamb2020time}, namely, the independence of the underlying system and intrinsic linearity. Based on these properties, our proposed model uses a finite-dimensional approximation of the linear Koopman operator to propagate the convolutional coordinates of SPOD in time.

\begin{figure*}[ht!]
		\centering
        \includegraphics[width=1\textwidth]{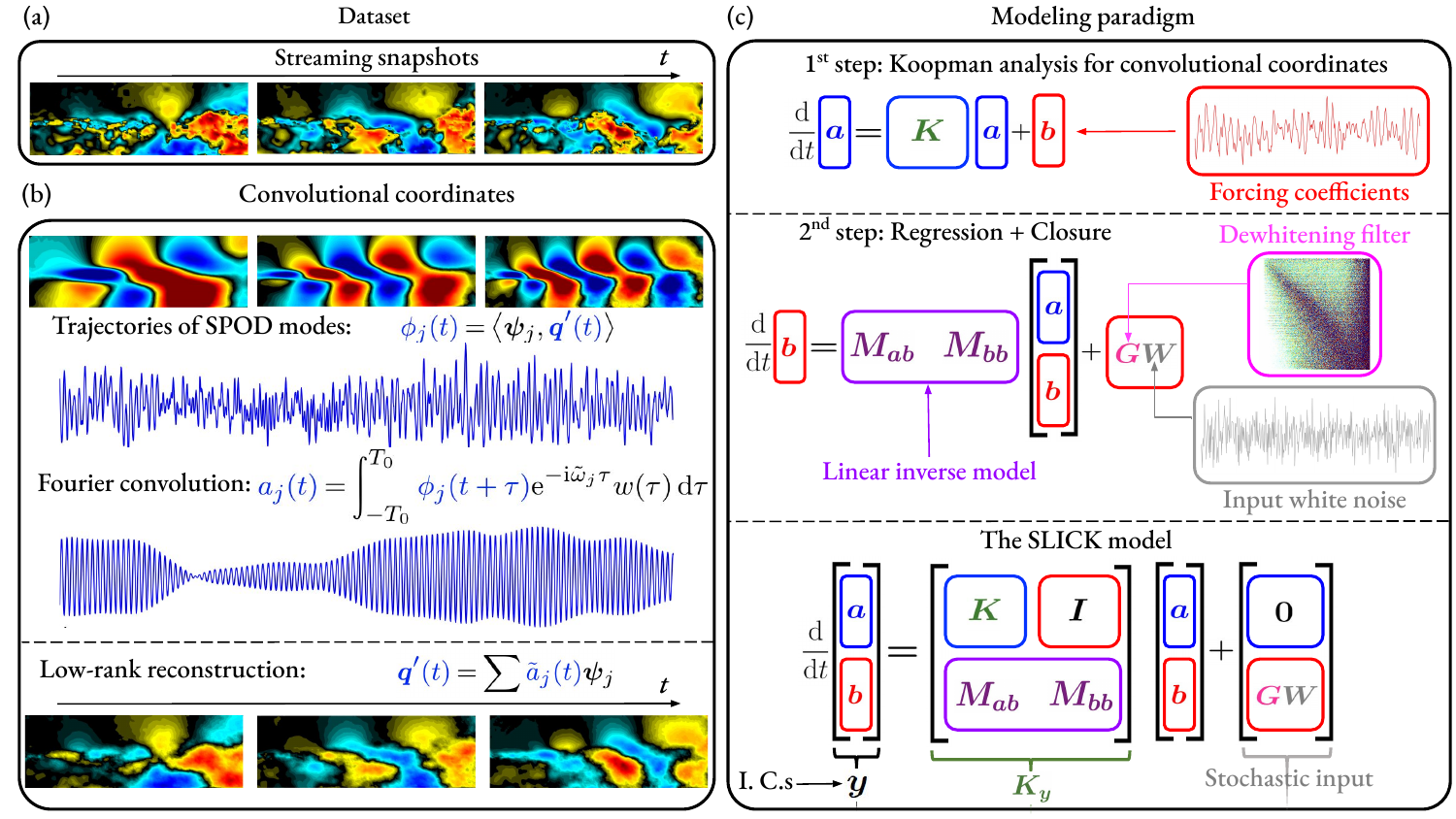} 
        \caption{Schematic of the SLICK model for broadband turbulent flows: (a) collecting data as equally sampled snapshots; (b) performing model order reduction and Fourier convolution; (c) training the SLICK model. The open cavity flow later presented in \S \ref{cavity} is shown as an example.
        }  \label{Koopman_SPOD_schematic}
	\end{figure*}
 

For fully turbulent flows, it is simply not realistic, at least not in a ROM spirit, to have a linear operator $\vb*{K}$ governing the dynamics of all temporal scales.
In our model, $\vb*{K}$ exclusively propagates the most energetic coherent structures obtained through SPOD,
whose time-coherence facilitates model prediction.
Nonlinear interactions that are not modeled by $\vb*{K}$ and the residual broadband turbulence are subsequently modeled as a time-invariant forcing, $\vb*{b}$.
Turbulence closure models that model $\vb*{b}$ directly as stochastic forcing have been proposed in the past, see, e.g., \cite{majda2012physics, sapsis2013statistically, majda2014conceptual}.
These approaches, however, neglect non-stochastic nonlinear interactions.
To model the nonlinear dynamics still contained in $\vb*{b}$, we inflate the model in equation (\ref{koopman_spod}) to incorporate $\vb*{b}$ into the state, analogous to EDMD that augments DMD with nonlinear observables, in the form of 
\begin{align}\label{2_lvl_model}
    \dv{}{t}\mqty[\vb*{a}\\ \vb*{b} ]=
    \left[
\begin{array}{c c}
 \vb*{K} & \vb*{I} \\
    \multicolumn{2}{c}{   \mathrel{\vcenter{\hbox{\rule{0.3cm}{0.6pt}}}}\vb*{M} \mathrel{\vcenter{\hbox{\rule{0.3cm}{0.6pt}}}}}
\end{array}
\right]
\mqty[\vb*{a}\\ \vb*{b} ]+ \mqty[\vb*{0}\\ \vb*{Gw} ]. 
\end{align}
Here, the dynamics of the forcing, $\dv{}{t}\vb*{b}$, are separated into two parts.
The first part is a deterministic linear function of the current state and the forcing, modeled by the rectangle matrix $\vb*{M}$.
The second part pertains to turbulence closure through a stochastic source $\vb*{Gw}$. The stochastic source comprises Gaussian white noise $\vb*{w}$ and is colored by a dewhitening filter $\vb*{G}$ obtained from the data. It guarantees that the model accurately reproduces the second-order flow statistics and establishes the inter-frequencies dependence of the state. 
The resulting stochastic low-dimensional inflated convolutional Koopman (SLICK) model is a stochastically forced linear system that 
 drives the evolution of the compound time-delay coefficients and
has the form of a multi-level regression (MLR) model \cite{kravtsov2005multilevel,kondrashov2005hierarchy} truncated at the second level.
Following the linear inverse modeling (LIM) paradigm by Penland \cite{penland1989random}, $\vb*{b}$ is informed by 
the offset between the linear dynamics and the true state evolution in the training phase of the model via linear regression.
Unlike the operator-based model by Chu $\&$ Schmidt \cite{chu2021stochastic}, the SLICK model is purely data-driven and does not require the knowledge of the linearized Navier-Stokes operator.
In fact, it can be applied to stationary discrete time-continuous data from any physical scenario.
Figure~\ref{Koopman_SPOD_schematic} shows a graphical summary of the SLICK model using the open cavity flow later in \S \ref{cavity} as an example.

 This paper is organized as follows. 
 In \S \ref{Koopman}, we introduce Koopman theory with time-delay embedding and demonstrate the optimality of SPOD modes as the modal basis when Fourier convolution is employed. 
 \S \ref{2-lvl model} introduces the SLICK model.
The performance of the model is demonstrated in \S \ref{results} across three benchmark problems: stochastic complex Ginzburg–Landau equation (SCGL), large eddy simulation (LES) data of a high Reynolds number turbulent jet, and particle image velocity (PIV) data of a high Reynolds number open cavity flow.
 Finally, \S \ref{conclusion} concludes and summarizes the paper.

\section{Theoretical Background} \label{Koopman}

As a theoretical means to obtain the evolution operator for the flow state, we first introduce Koopman analysis in \S \ref{Koopman_def}.
Since we are using linear observables for Koopman analysis, we require time-delay embedding to capture nonlinear effects. This is described in \S \ref{time_delay}.
The optimal convolutional coordinates that realize this time-delay embedding are provided by SPOD, as we will show in \S \ref{SPOD}.
The leading SPOD modes as a modal basis only describe the motion of large coherent structures. 
The unresolved scales, that is, the background turbulence, are not represented by this basis and have yet to be modeled.
To this end, we augment the linear model with an exogenous forcing in \S \ref{SPOD-Koopman}, which results in the final form of a linear time-invariant system.


\subsection{Koopman analysis}\label{Koopman_def}

The fundamental concept of Koopman analysis \cite{koopman1931hamiltonian} is to use an infinite-dimensional linear operator $\mathcal{K}$ that advances observable functions $\phi$ forward in time to describe the finite-dimensional nonlinear dynamics of the flow state $\vb*{q}\in \mathcal{M}$, where $\mathcal{M}\subseteq \mathbb{C}^{N_q}$ is the state space.
Let $\vb*{F}:\, \mathcal{M} \to \mathcal{M}$ be an
evolution operator that describes the nonlinear dynamics of $\vb*{q}$ with
\begin{align}
\vb*{q}{[i+1]} = \vb*{F}(\vb*{q}{[i]}).
\end{align}
The Koopman operator acts on functions of state space and is defined by
\begin{align}\label{koopman}
 \mathcal{K} \vb*{\phi}(\vb*{q}[i]) \equiv \vb*{\phi}(\vb*{F}(\vb*{q}[i]))= \vb*{\phi}(\vb*{q} [i+1]),
\end{align}
producing a new vector-valued function $\vb*{\phi} \circ \vb*{F}$.
Here, the observable vector $\vb*{\phi}$ is defined as
\begin{align} \label{g_vector}
    \vb*{\phi}(\vb*{q}) =\mqty[ \phi_1(\vb*{q}) &  \phi_2(\vb*{q})& \cdots & \phi_{N_k}(\vb*{q})]^T,
\end{align}
where $(\cdot)^T$ denotes the transpose.
Each component of $\vb*{\phi}$ is a scalar-valued observable with $\phi :\, \mathcal{M}\to \mathbb{C}$.  For intuitive understanding, we recommend the notes by Brunton \cite{Brunton2019NotesOK}.
The most important property of the Koopman operator is that it linearly governs the evolution of observables in discrete time, i.e., $\mathcal{K}(c_1 \vb*{\phi}_1 +c_2\vb*{\phi}_2) = c_1 \mathcal{K} \vb*{\phi}_1 +c_2 \mathcal{K} \vb*{\phi}_2$.
We omit the dependence on $\vb*{q}$, and use the notation $\vb*{\phi}[i]\equiv \vb*{\phi}(\vb*{q}[i])$ for brevity.
In general, the Koopman operator can be approximated by a finite-dimensional matrix $\vb*{K}_\text{EDMD} \in \mathbb{C}^{N_k\times N_k}$ obtained using EDMD \citep{williams2015data}. EDMD converges to a Galerkin method in the large-data limit \citep{williams2015data}, $N\to \infty$.
In this limit, the EDMD operator recovers the action of the Koopman operator on the $N_k-$dimensional subspace of observables \citep{klus2015numerical}, and the infinite-dimensional Koopman operator under the assumption of infinite observables, i.e., $N_k\to \infty$ \citep{korda2018convergence}.
In standard DMD \cite{schmid2010dynamic}, the full state observable vector is defined as the observable vector, that is, $\vb*{\phi}(\vb*{q})=\vb*{q}$.
For stationary flow, the physically relevant DMD modes often have nearly zero growth/decay rates, see, e.g., 
\cite{rowley2009spectral,schmid2010dynamic,schmid2012decomposition}.
This is not surprising as DMD modes become Fourier modes for periodic flows with zero-mean \cite{chen2012variants}.
It is commonly recommended not to subtract the mean when performing DMD for non-stationary flows to capture their dynamics.
If one aims to recover Koopman modes of stationary flows with zero growth/decay rates, incorporating mean subtraction in DMD is beneficial \cite{towne2018spectral}.



We formalize the mean flow subtraction by taking the Reynolds decomposition of the turbulent flow state, $\vb*{q}$, into the temporal mean, $\overline{(\cdot)}$, and fluctuating components, $(\cdot)'$, respectively, as
\begin{align}
    \vb*{q}[i]=\overline{\vb*{q}}+\vb*{q}'[i]. \label{flow_d}
\end{align}
We seek a dynamical system that describes the evolution of $\vb*{q}'$.
Model order reduction of turbulent flows intrinsically requires aggressive truncation to a limited number of observables, i.e., $N_k\ll N_q$.
The art of the Koopman theory lies in selecting the right observables, and it can embed the state dynamics into a low-dimensional manifold \citep{budivsic2012applied}.
The arguably simplest way to reduce dimensionality is to use the amplitudes of spatial modes obtained via projection as observables.
When the spatial modes, $\vb*{\psi}_j$, are orthogonal, the projection reduces to the inner product
\begin{align}\label{observable_inner}
    \phi_j(\vb*{q}) = \left<\vb*{\psi}_j,\vb*{q}-\overline{\vb*{q}}\right>_{x} = \left<\vb*{\psi}_j,\vb*{q}'\right>_{x},
\end{align}
where the spatial inner product $\left<\cdot,\cdot \right>$ and associated norm are defined as
\begin{align}\label{energy}
    \left<\vb*{q}_1,\vb*{q}_2\right>_{x} = \int_{\Omega} \vb*{q}_1^*(\vb*{x})\vb*{W}(\vb*{x})\vb*{q}_2(\vb*{x}) \mathrm{d}\,\vb*{x}, \quad \text{and} \quad 
    \norm{\vb*{q}}^2_{x} = \left<\vb*{q},\vb*{q}\right>_{x},
\end{align}
respectively.
Here, $\vb*{W}$ is a diagonal positive-definite weight matrix, and $(\cdot)^*$ denotes the Hermitian transpose. We leverage the cheap-to-compute property of the inner product to obtain observables without requiring orthogonality of the spatial modes. The choice of the modal basis, 
$\{\vb*{\psi}_j\}_{j=1}^{N_k}$, hasn't been specified yet and will be discussed later in \S \ref{SPOD}. 
The linear observable vector $\vb*{\phi}$, however, may be overly constrained for adequately describing the complex dynamics encountered in fluids or other nonlinear systems.
We then explore the application of time-delay embedding as a means to capture the nonlinear dynamics.



\subsection{Time-delay embedding} \label{time_delay}




 


Time-delay embedding has been incorporated with Koopman theory to represent chaotic systems as linear dynamic models, such as 
the Hankel-DMD model \cite{arbabi2017ergodic}, the HODMD model \citep{le2017hodmd}, and the HAVOK model \citep{brunton2017chaos}.
These approaches obtain time-delay observables by performing the SVD of a large Hankel matrix.
Alternatively, we may define a general time-delay observable vector $\vb*{a}\in \mathbb{C}^{N_k}$ as a linear combination of previous and future observables, $\vb*{\phi}$, with
\begin{align}\label{HO_observable}
    \vb*{a}[i] \equiv \sum_{h=1-N_f}^{N_f} \mathcal{C}_{h} \left( \vb*{\phi} [i+h] \right),
\end{align}
where the linear maps, $\mathcal{C}_{h}:\,\mathbb{C}^{N_k}\to \mathbb{C}^{N_k} $, 
assign weights for different instants. 
The linearity of the Koopman operator $\mathcal{K}$ yields
\begin{align}\label{HO_koopman}
    \vb*{a}[i+1]  =  \sum_{h=2-N_f}^{N_f+1} \mathcal{C}_{h} \left( \vb*{\phi} [i+h] \right) =  \sum_{h=1-N_f}^{N_f} \mathcal{C}_{h} \left( \mathcal{K} \vb*{\phi} [i+h] \right)   =  \mathcal{K} \left(\sum_{h=1-N_f}^{N_f}  \mathcal{C}_{h} \left( \vb*{\phi} [i+h] \right)\right),   
\end{align}
which gives
\begin{align}\label{koopman_a}
    \vb*{a}[i+1] =  \mathcal{K} \vb*{a}[i].
\end{align}
This implies that $\vb*{a}$ spans a Koopman-invariant subspace, that is, the same Koopman operator $\mathcal{K}$ governs the evolution of both the observables, $\vb*{\phi}$, and the corresponding time-delay observables, $\vb*{a}$.
Equation (\ref{HO_koopman}) also exhibits an inherent equivalence to HODMD. 
By comparing the second and the third terms and shifting the time index with $N_f$, $\vb*{\phi}[i+1]$
can be cast into a linear combination of the previous $2N_f$ states, that is,
\begin{align} \label{HODMD_infty}
    \vb*{\phi}[i+1] = & \quad \underbrace{\mathcal{C}_{N_f}^{-1}\left(\mathcal{K}\circ\mathcal{C}_{N_f}-\mathcal{C}_{N_f-1}\right)}_{ \mathcal{R}_{2N_f} } \vb*{\phi}[i]+ \underbrace{\mathcal{C}_{N_f}^{-1}\left(\mathcal{K}\circ\mathcal{C}_{N_f-1}-\mathcal{C}_{N_f-2}\right)}_{ \mathcal{R}_{2N_f-1} } \vb*{\phi}[i-1]+\cdots \\
    & + \underbrace{\mathcal{C}_{N_f}^{-1}\left(\mathcal{K}\circ\mathcal{C}_{2-N_f}-\mathcal{C}_{1-N_f}\right)}_{\mathcal{R}_{2}} \vb*{\phi}[i-2N_f+2] \nonumber    +\underbrace{\mathcal{C}_{N_f}^{-1}\left(\mathcal{K}\circ\mathcal{C}_{1-N_f}\right)}_{\mathcal{R}_{1}} \vb*{\phi}[i-2N_f+1],
\end{align}
where $\mathcal{C}^{-1}$ represents the inverse operator of $\mathcal{C}$. 
Here, the linear operators $\mathcal{R}_1,\cdots,\mathcal{R}_{2N_f}$ are constructed as linear combinations of the Koopman operator $\mathcal{K}$ and linear maps $\mathcal{C}$. Their finite-dimensional representations recover the formalization of HODMD.



We use equation (\ref{HO_observable}) to specialize the construction of the time-delay coordinates as a convolution process by assigning the linear maps as
   \begin{align}
    \mathcal{C}_h(\vb*{\phi})  \equiv \mqty[ c_{h,1}\phi_1 &  c_{h,2}\phi_2& \cdots & c_{h,N_k}\phi_{N_k}]^T,
\end{align}
where the $c_{h,1},\cdots,c_{h,N_k}$ are nonzero convolutional weights. Together with equation (\ref{HO_koopman}), we formalize the idea of using a Koopman operator to evolve the convolutional coordinates.
This idea has been previously explored by Kamb $\textit{et al.}$ \cite{kamb2020time}.
Most notably, as mentioned before, the Koopman operator remains invariant under the convolution.
Hankel singular vectors are particularly well-suited for convolution bases as they 
converge to Legendre polynomials in the limit of short delays \citep{gibson1992analytic} and to a Fourier basis in the limit of long delays \citep{bozzo2010relationship}.
The natural choice for the convolution basis for a statistically stationary process is the Fourier basis.
We avoid the construction of a Hankel matrix altogether by considering the Fourier convolution,
\begin{align} \label{conv_coeff}
    {a}_j(t) \equiv  \int_{-T_0} ^{T_0}  \phi_j{(t+\tau)} \mathrm{e}^{-\mathrm{i}\tilde{\omega}_{j} \tau}w(\tau) \, \dd \tau.
\end{align}
to obtain the convolutional coordinates, $a_j$, that describe the temporal dynamics of individual frequency components. Here, $\tilde{\omega}_{j}\in\{\omega_l\}_{l=1}^{2N_f}$ is the associated frequency. The Fourier series, $\{\mathrm{e}^{-\mathrm{i}\omega_l \tau}\}_{l=1}^{2N_f}$, are orthonormal on the time interval $[-T_0, T_0]$.
In practice, a window function $w(\tau)$ is included to minimize spectral leakage.
The discrete form of equation (\ref{conv_coeff}) on the time interval $(-N_f \Delta t,N_f \Delta t]$ can be written as
\begin{align} \label{conv_observable}
    {a}_j[i]=  \sum_{h=1-N_f}^{N_f} c_{h,j} \phi_j[i+h] 
\end{align}
by defining the convolutional weights as
\begin{align}\label{conv_w}
    c_{h,j} \equiv \mathrm{e}^{-\mathrm{i}\tilde{\omega}_j h\Delta t}w[h\Delta t]\Delta t.
\end{align}
Equation (\ref{conv_observable}) is a continuously discrete convolution sum for the observable $\phi$. Instead of explicitly computing the convolutional weights, we use the fast Fourier transform (FFT) to perform the discrete convolution.
We next demonstrate that SPOD provides the optimal modal basis for the Fourier convolution.













    

\subsection{Optimal convolutional coordinates are SPOD coefficients} \label{SPOD}

The first step to finding the modal basis vector is to obtain the span of the Fourier basis.
 We decompose the fluctuating state into its temporal discrete Fourier modes, $\hat{\left(\cdot\right)}$, as
\begin{align}\label{q_DFT}
    \vb*{q}'(t) = \frac{1}{2N_f\Delta t}\sum_{l=1}^{2N_f} \hat{\vb*{q}}(\omega_l) \mathrm{e}^{\mathrm{i} \omega_l t}
\end{align}
on the time interval $(-N_f \Delta t,N_f \Delta t]$,
where $\omega_l =  l \pi /(N_f \Delta t)$ is the angular frequency. 
Inserting equation (\ref{observable_inner}) into equation (\ref{q_DFT}) yields the $j$th observable
\begin{align} \label{phi_j}
    \phi_j(\vb*{q}) = \frac{1}{2N_f\Delta t}\sum_{l=1}^{2N_f}\left<\vb*{\psi}_j,\hat{\vb*{q}} (\omega_l)\right>_x \mathrm{e}^{\mathrm{i} \omega_l t}.
\end{align}
Substituting this expression for the observable in equation (\ref{conv_observable}) and using the definition of the convolutional weights, $c_{h,j}$, from equation (\ref{conv_w}) leads to
\begin{align}\label{a_j}
    {a}_j[i]  = \sum_{h=1-N_f}^{N_f} c_{h,j} \left( \frac{1}{2N_f\Delta t}\sum_{l=1}^{2N_f}\left<\vb*{\psi}_j,\hat{\vb*{q}} (\omega_l)\right>_x \mathrm{e}^{\mathrm{i} \omega_l (i+h)\Delta t} \right) =   \left<\vb*{\psi}_{j},\hat{\vb*{q}}(\tilde{\omega}_j)\right>_x \mathrm{e}^{\mathrm{i} \tilde{\omega}_j (i \Delta t)}.
\end{align}
The magnitude of $a_j$ only depends on the inner product of the $j$th basis vector, $\vb*{\psi}_{j}$, and the Fourier mode associated with frequency $\tilde{\omega}_j$, $\hat{\vb*{q}}(\tilde{\omega}_j)$.
With modes that have unit energy, $ {\|\vb*{\psi}_{j}\|_x=1}$, we seek the convolutional coordinates that optimally represent the flow field at each given frequency, $\tilde{\omega}_j$. This objective is formalized by maximizing
the quantity
\begin{align}
    \lambda_j = 
    \mathrm{E}\{ a_ja_j^*\}  = \left< \vb*{\psi}_{j}, \vb*{S}_j \vb*{W} \vb*{\psi}_{j}   \right>_x, \label{Rayleigh_quotient}
\end{align}
 where $\mathrm{E}\{\cdot\}$ denotes the expected value over a large number of realizations, and $\vb*{S}_j=\mathrm{E}\{ \hat{\vb*{q}}(\tilde{\omega}_j) \hat{\vb*{q}}(\tilde{\omega}_j)^* \}$ is the cross-spectral density matrix. The detailed derivation of equations (\ref{a_j}-\ref{Rayleigh_quotient}) is provided in the Supplementary Material.
 Equation (\ref{Rayleigh_quotient}) represents a generalized Rayleigh quotient, and its maximum values can be found from the eigenvalue problem
 \begin{align}\label{spod_cros}
      \vb*{S}_j \vb*{W} \vb*{\psi}_{j}^{(\alpha)}  = \lambda_j^{(\alpha)} \vb*{\psi}_{j}^{(\alpha)}.
 \end{align}
 The corresponding eigenvectors, $\vb*{\psi}_{j}^{(\alpha)}$, are known as SPOD modes. Refer to Towne \textit{et al.}\cite{towne2018spectral} for more details. At the same frequency, these modes are orthogonal to each other, i.e., $\left<\vb*{\psi}_j^{(\alpha)}, \vb*{\psi}_j^{(\beta)}\right>_x = \delta(\alpha - \beta).$
Even though the SPOD modes across different frequencies are not orthogonal in the spatial inner product, $\left<\cdot , \cdot\right>_x$, the convolution in equation (\ref{a_j}) reinstates the space-time orthogonality, $\int_{-T_0}^{T_0}\left<\vb*{\psi}_j^{(\alpha)} \mathrm{e}^{\mathrm{i} \tilde{\omega}_j t}, \vb*{\psi}_k^{(\beta)}\mathrm{e}^{\mathrm{i} \tilde{\omega}_k t}\right>_x\, \dd t = 2T_0 \delta(\alpha - \beta)\delta(\tilde{\omega}_j-\tilde{\omega}_k)$, thereby preserving the direct correspondence between modes and frequencies inherent to SPOD.
The optimality we require for the convolutional Koopman model hence naturally recovers SPOD modes as the adequate basis. 
To represent the dynamics at all scales of turbulence, we choose a modal basis that includes all $2N_f$ frequency components.
At each frequency, we pick the leading $M$ modes to retain the largest amount of energy.
The resulting basis takes the form of 
{{
\begin{equation}
  \vb*{V}= \left[
  \smash[b]{\underbrace{\begin{matrix}
  \vert &  &\vert \\
 \vb*{\psi}_{1}^{(1)} & \cdots &\vb*{\psi}_{1}^{(M)}\\
  \vert & &\vert
  \end{matrix}}_{\omega_1 }}  \quad  \cdots \quad 
   \smash[b]{\underbrace{\begin{matrix}
  \vert & &\vert \\
  \vb*{\psi}_{2N_f}^{(1)} & \cdots &\vb*{\psi}_{2N_f}^{(M)}\\
  \vert & &\vert
  \end{matrix}}_{\omega_{2N_f} }}
  \right], \label{V}
\end{equation}} \vskip 0ex}
{\flushleft{
which yields $N_k=2MN_f$ observables. We denote it as a rank $M\times 2N_f$ SPOD basis. For real data, only $N_k = M(N_f+1)$ non-negative frequency components are needed to}} form the basis.
The $j$th column of $\vb*{V}$ is the $(j-\lfloor j/M\rfloor M)$th SPOD mode at the frequency $\tilde{\omega}_j = \omega_{\lceil j/M\rceil}$.
Inserting equation (\ref{phi_j}) into the Fourier convolution (\ref{conv_observable})
recovers the convolutional integral to compute discrete time-continuous SPOD expansion coefficients \citep{nekkanti2021frequency}. 
This implies that the convolutional coordinates, $\vb*{a}$, facilitate the optimal low-reconstruction of the flow field in the space-time sense, $\int_{-T_0}^{T_0}\left<\vb*{q}_1 , \vb*{q}_2\right>_x\, \dd t $. Details for the reconstruction are provided in the Supplementary Material.

\subsection{Forced LTI system} \label{SPOD-Koopman}





The use of convolutional coordinates implicitly embeds nonlinearity into the Koopman operator, $\mathcal{K}$.
For fully turbulent flows, it is unrealistic to expect a truncated linear operator, $\vb*{K}$, to govern the dynamics of all temporal scales.
We hence suggest augmenting the linear Koopman model in equation (\ref{koopman_a}) with an exogenous time-invariant forcing, $\vb*{b}$, to account for the remaining nonlinear interactions and background turbulence.
The resulting evolution model for the discrete time-continuous convolutional coordinates can be written as a continuous-time LTI system,
\begin{align} 
\dv{}{t}{\vb*{a}} = 
    \vb*{K} \vb*{a}+\vb*{b}. \label{koopman_cont}
\end{align}  
Here, $\vb*{K}\in \mathbb{C}^{N_k\times N_k}$ is a finite-dimensional approximation of the continuous-time version of the Koopman operator. 
In the absence of $\vb*{b}$, the matrix
$\vb*{K}$ is learned from the empirical convolutional coordinates, $\check{\vb*{a}}$, obtained through the SPOD of the data, $\check{\vb*{q}}$.
The matrix $\vb*{K}$ is then obtained
by solving the least-squares problem with $L_2$ regularization \cite{hoerl1970ridge},
\begin{align}  \label{K_a}
    \vb*{K}=\arg\min_{\vb*{K}} \frac{1}{2} \left( \sum_{i=1}^{N-1}\norm{\dv{}{t}\check{\vb*{a}}[i+1]-\vb*{K}\check{\vb*{a}}[i] }^2 +\gamma_1 \norm{\vb*{K} }^2 \right),
\end{align}
where $\gamma_1$ is the ridge parameter. 
$L_2$ regularization is particularly well-suited for enhancing model stability when the training data is noisy, which is true for turbulent flows.
Instead of considering $\vb*{b}$ as merely stochastic, we treat $\vb*{b}$ as an independent dynamical variable that models the residual nonlinear dynamics. Our approach is next discussed in \S \ref{2-lvl model}.




\section{The stochastic low-dimensional inflated convolutional Koopman (SLICK) model} \label{2-lvl model}






The core concept of Koopman theory is to use an infinite-dimensional linear operator that acts on observables to describe nonlinear dynamics.
Incorporating nonlinear observables, in addition to linear observables, is necessary to fully characterize the behavior of a nonlinear system.
EDMD \cite{williams2015data} follows this idea and inflates the standard DMD with additional nonlinear observables to provide a better finite-dimensional approximation of the Koopman operator.
The MLR models \cite{kondrashov2005hierarchy,kravtsov2005multilevel} introduced in \S \ref{Sec:intro} can hence be interpreted as a Koopman perspective: each level of the linear model inflates the state vector by the residue of the previous state. 
Along these lines, we inflate the LTI system (\ref{koopman_cont}) to incorporate the dynamics of $\vb*{b}$ within the model, which yields
\begin{align}
    \dv{}{t} \underbrace{\mqty[\vb*{a}\\ \vb*{b} ]}_{\vb*{y}}=
    \underbrace{\left[
\begin{array}{c c}
  \vb*{K} & \vb*{I} \\
 \multicolumn{2}{c}{   \mathrel{\vcenter{\hbox{\rule{0.3cm}{0.6pt}}}}\vb*{M} \mathrel{\vcenter{\hbox{\rule{0.3cm}{0.6pt}}}}}
\end{array}
\right]}_{\vb*{K}_y}
\mqty[\vb*{a}\\ \vb*{b} ]+ {\mqty[\vb*{0}\\ \vb*{r} ]}.  \label{dydt_0}
\end{align}
Here, the square matrix, $\vb*{K}$, drives the motion of the most energetic coherent structures, represented by $\vb*{a}$. Additionally, the rectangular matrix, $\vb*{M}$, governs the evolution of $\vb*{b}$ 
and slaves it to the state $\vb*{a}$. The compound matrix, $\vb*{K}_y$, governs the linear dynamics of the inflated convolutional coordinates, $\vb*{y}$. Therefore, it can be interpreted as a finite-dimensional approximation of the Koopman operator within the time-delay framework. 
Due to this finite dimensionality, the model--for non-trivial flows--necessarily has a residual error $\vb*{r}$.
With the chaotic nature of turbulent flows in mind and the goal of conducting Monte Carlo-based uncertainty quantification, we propose a stochastic closure model. In this closure, the residue is modeled as a stochastic source,
 \begin{align} \label{residue}
     \vb*{r} \approx \vb*{G w}.
 \end{align}
Here, white noise input, $\vb*{w}$, is colored by a dewhitening filter, $\vb*{G}$, that is informed by the data.
This form of closure truncates the MLR models at the second-level.
The goal of using $\vb*{G}$ is to preserve the same second-order flow statistics.
In the remainder of this section, we will discuss the identification of the matrices $\vb*{M}$ and $\vb*{G}$ from the data.

To incorporate $\vb*{b}$ within the model, the key idea is to use the offset between the linearized dynamics and the training data,
\begin{subequations}
  \begin{align}
  \check{\vb*{b}}&= \dv{}{t}\check{\vb*{a}} - \vb*{K} \check{\vb*{a}},
\end{align}  
\end{subequations} 
to inform the model in the training phase via linear regression.
Again, the same empirical convolutional coordinates, $\vb*{\check{a}}$, employed in equation (\ref{K_a}) to derive $\vb*{K}$ are utilized for training.
This follows the inverse modeling paradigm \cite{penland1989random,kondrashov2005hierarchy,kravtsov2005multilevel}.
In practice, we determine the still-existing nonlinear dynamics in $\vb*{b}$ by 
solving 
\begin{align}\label{M}
   \vb*{M} \equiv \arg\min_{\vb*{M}}\sum_{i=1}^N\left( \norm{\dv{}{t}\check{\vb*{b}}[i]-\vb*{M}\check{\vb*{y}}[i] }^2+\gamma_2 \|\vb*{M}\|^2\right)
\end{align}
with a ridge parameter $\gamma_2$, where $\vb*{\check{y}}$ represents the empirical inflated states containing $\vb*{\check{a}}$ and $\vb*{\check{b}}$.


Given an initial condition and the stochastic input, equation (\ref{dydt_0}) can be integrated numerically using the Euler–Maruyama method as
\begin{align}
      \vb*{y}[i+1] = {\vb*{T}}\vb*{y}[i] +
      \mqty[\vb*{0} \\\vb*{G } \Delta \vb*{\xi}], \quad \text{where} \quad \vb*{T}\equiv \vb*{I}+\vb*{K}_{y}\Delta t. \label{sde_d}
\end{align}
Here, $\Delta \vb*{\xi}$ is an independent Gaussian increment of a Wiener process with zero mean and covariance matrix $\vb*{I}\Delta t$. Equation (\ref{sde_d}) is the discrete version of equation (\ref{dydt_0}), where $\Delta \vb*{\xi}$
substitutes $\vb*{w}$ in equation (\ref{residue}) through integration.
By taking the expectation of equation (\ref{sde_d}), 
the equation that propagates the expected value of $\vb*{y}[i]$ can be obtained as
\begin{align}
      \mathrm{E}\{\vb*{y}[i+1]\} = {\vb*{T}} \mathrm{E}\{\vb*{y}[i]\}. \label{sde_d_exp}
\end{align}
The remaining task is to construct the dewhitening filter, $\vb*{G}$, to ensure accurate reproduction of the second-order flow statistics by the model---that is, the covariance matrix of the inflated state, $\vb*{P}[i]\equiv\mathrm{E}\qty\Big{\left(\vb*{y}[i]- \mathrm{E}\{\vb*{y}[i]\}\right) \left(\vb*{y}[i]- \mathrm{E}\{\vb*{y}[i]\}\right)^{^*}}$.
The evolution equation for the latter is readily obtained by subtracting equation (\ref{sde_d_exp}) from equation (\ref{sde_d}) and then right-multiplying by its Hermitian transpose. This yields
\begin{align}\label{y_P_prop}
    \vb*{P}[i+1] = \vb*{T}\vb*{P}[i]\vb*{T}^*+\mqty[ \vb*{0} & \vb*{0}\\
    \vb*{0}& \vb*{G G}^*\Delta t].
\end{align}
If $\vb*{T}$ is stable, $\lim_{i\to \infty} \mathrm{E} \{\vb*{P}[i]\}$ converges to the solution of a discrete-time Lyapunov equation, 
\begin{align}\label{lyap}
    \vb*{T P T}^*-\vb*{P}+\mqty[ \vb*{0} & \vb*{0}\\
    \vb*{0}& \vb*{G G}^*\Delta t]=0.
\end{align}
From equation (\ref{sde_d_exp}), its also clear that $\lim_{i\to \infty} \mathrm{E} \{\vb*{y}[i]\}$ converges to zero for stable eigenvalues. This provides an analytical uncertainty quantification (UQ) and guarantees that $\vb*{y}$ is bounded.

For finite data, we inform the dewhitening filter, $\vb*{G}$, directly from equation (\ref{y_P_prop}) by defining the shifted auto-covariance matrices,
\begin{align}
     \check{\vb*{P}}_1 \equiv \mathrm{E}\{ \left(\check{\vb*{y}}\check{\vb*{y}}^* \right)\vert_{i =1}^{N-1} \} \, \quad \text{and}\,\quad  \check{\vb*{P}}_{2} \equiv \mathrm{E}\{ \left(\check{\vb*{y}}\check{\vb*{y}}^* \right)\vert_{i= 2}^{N} \}.
\end{align}
from the training data. Substituting $\check{\vb*{P}}_1$ and $\check{\vb*{P}}_2$ into equation (\ref{y_P_prop}) allows us to solve for $\vb*{G G}^*$ as
\begin{align}\label{GG}
     \vb*{G G}^* \approx \vb*{H}= \frac{1}{\Delta t} \mqty[ \vb*{0} & 
    \vb*{1}] \underbrace{\left(  \check{\vb*{P}}_2 - \vb*{T}   \check{\vb*{P}}_1 \vb*{T}^*\right)}_{\check{\vb*{H}}} \mqty[ \vb*{0} \\ \vb*{1}].
\end{align}
Implicitly involved here is the ergodicity hypothesis. In the absence of true ensemble statistics for stochastic realizations evolving from the same initial condition under different stochastic inputs, we substitute $\vb*{P}[i]$ and $\vb*{P}[i+1]$ with the finite-time statistics of the training data. By construction, the matrix $\vb*{H}$ is Hermitian but not guaranteed positive semi-definite.
To obtain the closest positive semi-definite matrix of $\vb*{H}$ in the Frobenius norm \citep{higham1988computing}, we first perform the eigenvalue decomposition of $\vb*{H}$,
\begin{align}\label{H_EVD}
\vb*{H} = \vb*{V}_H \vb*{D}_H \vb*{V}_H^*,
\end{align}
and set its negative eigenvalues to zero.
Following \cite{penland1996stochastic}, we furthermore rescale the positive eigenvalues to maintain the trace of $\vb*{H}$ such that the total variance of the stochastic forcing is preserved. The dewhitening filter $\vb*{G}$ can then be determined as 
\begin{align}
      \vb*{G} = \vb*{V}_H \tilde{\vb*{D}}_H^{1/2},   \label{G}
\end{align}
where $\tilde{\vb*{D}}_H$ is the modified diagnal eigenvalue matrix. 
While the above stochastic low-dimensional inflated convolutional Koopman (SLICK) model encompasses complex concepts, it can be formulated in a simple manner, as illustrated in the schematic (figure~\ref{Koopman_SPOD_schematic}) and detailed in the discrete-time algorithm provided in the Supplementary Material.




 



\section{Examples} \label{results}

\FloatBarrier

\begin{table}[ht!]
\centering
\caption{Overview of datasets and modeling parameters. All quantities are described in the text.}\label{database}
\begin{tabular}{cccccccc}
 \multicolumn{7}{c}{Datasets} \\\hline
Flow  & Method	& Variables & $N$ & $N_q$  & Section & Overview \\\hline
SCGL & SRK4 \citep{kasdin1995runge} & $q$ & 80000   & 220  & \S \ref{G_L}  & Fig. \ref{SCGL_nonlinear} \\
 Jet  & LES \citep{bresetal_2018jfm} & $p$  & 10000 & $950{\times} 195$   & \S \ref{jet}  & Fig. \ref{jet_nfft_128}\\
Cavity   & TR-PIV \citep{zhang2017identification} & $u,\,v$  & 16000 & $156{\times} 55{\times}2$ & \S \ref{cavity} & Fig. \ref{cavity_nfft_256} \\\hline
\multicolumn{7}{c}{} \\
\multicolumn{7}{c}{Modeling parameters} \\\hline
Flow  	& $N_\text{train}$ & Delay  & Model rank & Energy  & $\gamma_1/\mathrm{E}\{\|\vb*{a}\|^2\}$ & $\gamma_2/\mathrm{E}\{\|\vb*{y}\|^2\}$  \\\hline
SCGL    & 20000  & 32   & 2$\times$32 & $92.5\%$ & 0 & 0\\
 Jet    & 9000 & $128$    & $15{\times}65$ & $83.6 \%$ & $10^{-3} $ & $10^{-4}$  \\
Cavity    & 14400& $256$   & $20{\times}129$ & $76.6\%$ & $10^{-3}$ & $10^{-4}$    \\\hline
\end{tabular}
\end{table}


We demonstrate the SLICK model using three different examples: a numerically integrated stochastic complex Ginzburg–Landau equation (SCGL) with spatially-correlated Gaussian white noise; the pressure component of a high-fidelity turbulent jet simulation by Brès $\textit{et al.}$ \cite{bresetal_2018jfm}; and experimental particle image open cavity flow data by Zhang \textit{et al.} \cite{zhang2017identification}, as summarized in Table \ref{database}. These examples are not just diverse sources of numerical data; the latter two are fundamental, large research-grade, state-of-the-art fluid mechanics datasets produced for previous physical exploration studies, see, e.g., \cite{schmidt2017wavepackets,schmidt2018spectral,nekkanti2021frequency} for the jet data and \cite{zhang2020spectral} for the cavity data.
Additionally, they are also representative of statistically stationary turbulent flows and exhibit different spectral content: the open cavity data has tonal peaks and an underlying broadband spectrum \citep{bresetal_2018jfm};
the turbulent jet data has broadband turbulence spectra \citep{zhang2020spectral}. 
The two research-grade data, in particular, underlie the need for data-driven modeling approaches: large high-fidelity numerical data become increasingly intractable for projection-based approaches; for experimental data, only an incomplete state consisting of two velocity components in a 2D plane is available. 
The following analysis focuses on three aspects of the model: short-to-medium prediction, prediction of long-term statistics, and uncertainty quantification. Beyond qualitative evaluation, two metrics, the Pearson correlation coefficient
\begin{align}
    \rho(\vb*{q}_1,\vb*{q}_2) =\frac{\mathrm{E}\qty\big{\left(\vb*{q}_1-\mathrm{E}\{\vb*{q}_1\}\right)^*\vb*{W}\left(\vb*{q}_2-\mathrm{E}\{\vb*{q}_2\}\right)}}{\sqrt{\mathrm{E}\qty\big{ \norm{ \vb*{q}_1-\mathrm{E}\{\vb*{q}_1\}}^2   }  \mathrm{E}\qty\big{ \norm{ \vb*{q}_2-\mathrm{E}\{\vb*{q}_2\}}^2   }} },
\end{align}
and the normalized root-mean-square error (RMSE)
\begin{align}
    \mathrm{RMSE}(\vb*{q}_1, \vb*{q}_2) = \sqrt{ \mathrm{E}\qty\Big{\frac{\|\vb*{q}_1-\vb*{q}_2\|^2}{\|\vb*{q}_2\|^2}} },
\end{align}
are employed to quantify the performance of the model. We consider the criteria with $\rho\in [0.9,1]$ for very strong correlation, $\rho\in [0.7,0.9)$ for strong correlation, $\rho\in [0.4,0.7)$ for moderate correlation, and $\rho\in [0.1,0.4)$ for weak correlation \citep{schober2018correlation}. We split the two research-grade data sets, characterized by challenges in acquiring a substantial number of snapshots, each comprising $N$ snapshots, into 90$\%$ training data and 10$\%$ test data, as shown in Table \ref{database}. 








\subsection{Stochastic complex Ginzburg–Landau equation} \label{G_L}
	\begin{figure*}
		\centering
        \includegraphics[trim = 0mm 0mm 0mm 0mm, width=.9\textwidth]{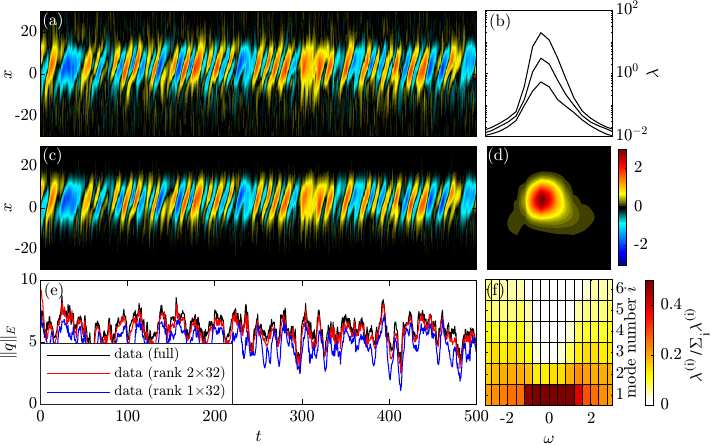}
        \caption{Overview of the stochastic complex Ginzburg-Landau data.
        Time domain representation in the left column shows temporal evolution for full data in (a), rank $2\times 32$ data in (c), and energy in (e). Frequency domain representation in the right column shows the first three SPOD eigenvalue spectra in (b), weighted leading SPOD modes $\sqrt{\overline{\|q\|}\lambda(\omega)} \psi(x;\omega)$ in (d), and the modal percentage energy accounted by each mode in (f) as functions of frequency.} \label{SCGL_nonlinear}
	\end{figure*}

We first consider the SCGL equation, widely used as a model to study instabilities in spatially evolving flows. However, we also note that the SCGL model is fundamentally different from the fluid mechanics cases in that its statistical convergence is extremely slow, but a large number of snapshots are readily available at an insignificant computational cost. 
We hence do not use the $90/10$ splitting ratio but instead compute the long data set for SPOD, as shown in Table \ref{database}. The convergence study is provided in the Supplementary Material.
The forecast ability of the model is demonstrated in the prediction of the dominant flow pattern.
The SCGL equation is given by 
\begin{align} \label{SGL}
    \pdv{q}{t}  =& \left(-\nu\pdv{}{x}+\gamma\pdv{^2}{x^2}+\mu\right){q} -\xi q \lvert q \rvert^2+ \int_{\Omega} g(x,x') \eta(x'){w(x')} \,\dd x',
\end{align}
where the parameter $\mu$ is expressed in a quadratic form as 
$
    \mu(x) = \mu_0-C_\mu^2+\frac{\mu_2}{2}x^2,
$
see, e.g., \cite{hunt1991instability,cossu1997global}.
The parameter selection follows Ilak $\textit{et al.}$ \cite{ilak2010model}. Specifically, we choose the supercritical case with $\mu_0=0.41$, where the linear component of the SCGL equation is globally unstable \citep{huerre2000open,chomaz2005global} and exhibits similarities to vortex shedding in the presence of crossflow past a cylinder \citep{provansal1987benard,huerre1990local}. Additionally, including a cubic nonlinearity with $\xi = 0.1$ induces a limit cycle behavior in the system.
The Gaussian white noise input, ${w}\sim \mathcal{N}(0,1)$, with a uniformly distributed phase, is spatially constrained with an exponential envelope of $\eta(x) = \mathrm{exp}[-(x/60)^{10}]$. 
The Gaussian kernel function,
$
g(x,x')= \frac{1}{4\sqrt{2\pi}}\mathrm{exp}\left[-\frac{1}{2}\left(\frac{x-x'}{4}\right)^2\right],
$
is utilized to spatially correlate the white noise input to enhance statistical convergence \citep{towne2018spectral}.
The computational domain, $x\in\left[-85,85\right]$, is discretized using $N_q=220$ uniformly distributed nodes. The fourth-order stochastic Runge-Kutta algorithm (SRK4) by Kasdin \cite{kasdin1995runge} is employed to numerically integrate equation (\ref{SGL}) over time.
Panels \ref{SCGL_nonlinear}(a,e) show the temporal evolution of the spatial pattern and the corresponding energy, respectively.
We perform SPOD using $N=20000$ snapshots at a time step of $\Delta t = 0.5$, a block size of $N_f=32$, and an overlap of $75\%$.
Panels \ref{SCGL_nonlinear}(b,d) show the first three SPOD eigenvalue spectra and the leading SPOD modes as a function of the angular frequency, $\omega = 2\pi f$, respectively.
The SPOD eigenvalue attains its peak value at $\omega=-0.39$.
Figure \ref{SCGL_nonlinear}(f) presents the energy distribution among different modes, indicating that the first two modes capture most of the energy, specifically accounting for $92.5\%$ of the total energy. As shown in panels \ref{SCGL_nonlinear}(c,e), the data reconstruction using $2\times32$ modes, $\check{\vb*{q}}_{2\times 32}$, is remarkably accurate. 
Hence, we will use the first two SPOD modes to construct the ROM. Regularization is not required for this case.

    \begin{figure}[ht]
		\centering
        \includegraphics[trim = 0mm 0mm 0mm 0mm, width=0.7\textwidth]{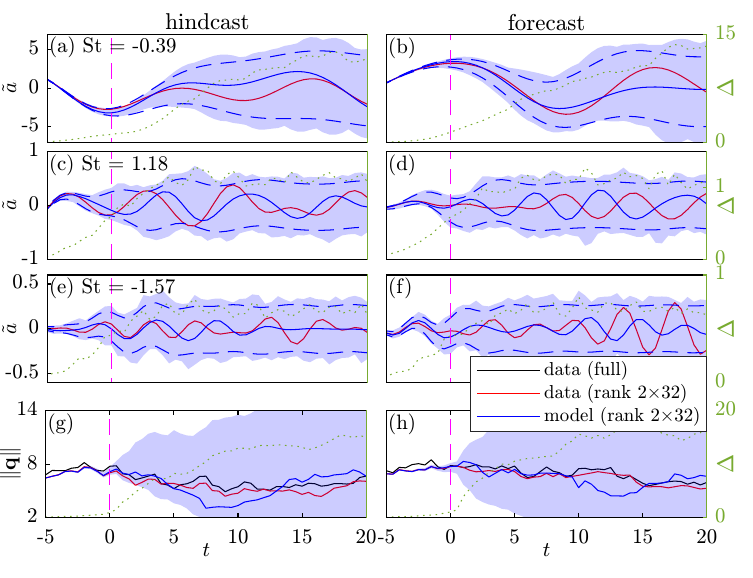}
        \caption{
        Hindcast (a,c,e,g) and forecast (b,d,f,h) for the SCGL equation 
in terms of the leading $\tilde{\vb*{a}}$ at frequencies: St=-0.39 (a,b), St=1.18 (c,d), and St=-1.57 (e,f), and the integral norm (g-h).
        A single realization of SLICK (blue) is compared to the rank 2$\times$32 data (red) and the full SCGL data (black).
        Magenta dashed lines show the onset of prediction. Blue dashed lines in (a-f) mark the analytic 95$\%$ confidence interval. The blue-shaded areas are Monte Carlo envelopes based on 5000 model evaluations. The widths of the uncertainty bound, $\Delta$, are shown as green dotted lines.
        } \label{SCGL_a_all}
	\end{figure}

\begin{figure}[ht]
		\centering
        \includegraphics[trim = 0mm 10mm 0mm 8mm, width=0.87\textwidth]{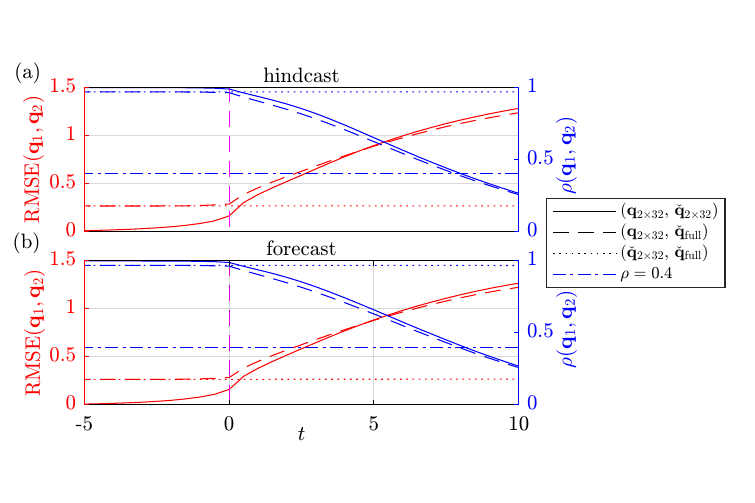}
        \caption{Normalized RMSE (left axes, red) and correlation (right axes, blue) for hindcast (a) and forecast (b). 
        Solid lines represent model--reduced-order data comparisons, dashed lines represent model--full data comparisons, and dotted lines represent reduced-order data--full data comparisons.
A moderate correlation of $\rho=0.4$ is highlighted as blue dashed lines.
For the SLICK model, 2000 initial conditions are each run with 200 Monte Carlo simulations. } \label{SCGL_error}
	\end{figure}

    We first assess the model using one initial condition within the training set. The left column of figure \ref{SCGL_a_all}, panels (a,c,e,g), compares the rank $2\times 32$ data and a single realization of the SLICK model for the hindcast. Starting from the onset of prediction ($t=0$), this example of realization follows the initial transient dynamics of the data for approximately 3-time units. Model UQ is performed using both the analytical 95\% confidence interval and 5000 Monte Carlo simulations; both metrics confirm that the model remains stability-bounded after an initial transient. This transient itself can be interpreted as the prediction horizon of the model, as discussed later in the context of figure \ref{SCGL_error}. Notably, the envelope of the uncertainty for total energy only tends to expand when the prediction starts. Results for forecasting are shown in the right column of figure \ref{SCGL_a_all}, panels (b,d,f,h). The overall trends closely resemble those obtained in the hindcast. This dataset serves as a model example due to its simplicity in equation integration. Abundant snapshots have been generated to well-converge statistics, resulting in nearly identical hindcast and forecast outcomes.


    \begin{figure}[ht]
		\centering
        \includegraphics[trim = 0mm 0mm 0mm 0mm, width=0.75\textwidth]{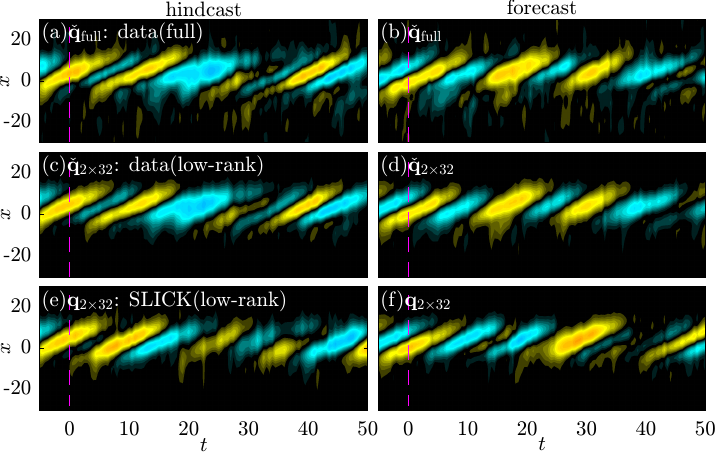}
        \caption{Spatial-temporal evolution of the full data (a-b), the rank $2\times 32$ data (c-d) and the rank $2\times 32$ SLICK (e-f) for hindcast (a,c,e) and forecast (b,d,f). } \label{SCGL_q_model}
	\end{figure}


After focusing on the temporal dynamics, we assess the prediction capability of SLICK for forecasting spatial structure. Figure \ref{SCGL_q_model} compares the spatial-temporal evolution of the full data, the rank $2\times 32$ data, and the model for both hindcast and forecast. The overall dynamics are, arguably, qualitatively well-predicted up to $t\approx 10$ for both.
After this period of very accurate forecasting, the model deviates from the full data while still predicting a highly realistic flow field.
Upon closer inspection of figure \ref{SCGL_a_all}, it becomes apparent that a combination of amplitude and phase errors is responsible for this deviation.



 Figure \ref{SCGL_error} shows the normalized RMSE and correlation for
2000 initial conditions. Each initial condition is used to run 200 Monte Carlo simulations.
The overall results for hindcast and forecast are comparable.
The model exhibits a slight deviation from the low-rank data at $t=0$, yet it demonstrates a nearly identical difference compared to the full data. We report the error here, but the analysis mainly focuses on correlation for consistent comparisons. The correlation coefficients are around 0.9 at $t\simeq 2$, 0.7 at $t\simeq 4.5$, and 0.4 at $t\simeq 8$. Therefore, we can consider the prediction horizon of the model for moderate positive correlation to be approximately $t=8$.






\begin{figure}[ht]
		\centering
        \includegraphics[trim = 0mm 0mm 0mm 0mm, width=0.35\textwidth]{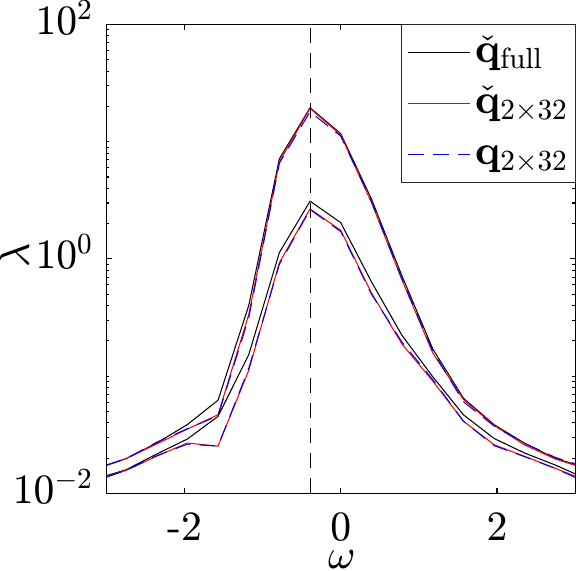}
        \caption{Comparison of SPOD eigenvalue spectra of the full data (black), rank $2\times32$ data (red), and rank $2\times32$ SLICK (blue).} \label{SCGL_spectra}
	\end{figure}

 
Starting from the initial condition, we expect the model to be predictive for $t\simeq 8$. Afterward, the model will run stably and produce a surrogate flow field, the statistics of which will be analyzed next. A natural choice is also to employ SPOD for this purpose.
 Figure \ref{SCGL_spectra} compares the leading two SPOD eigenvalues of $\check{\vb*{q}}_{\text{full}}$, $\check{\vb*{q}}_{2\times 32}$, and ${\vb*{q}}_{2\times 32}$. 
The SPOD eigenvalue spectra for the model and the low-rank data are indistinguishable. Both closely resemble those of the full data.
The aforementioned results validate that the rank $2\times32$ model accurately performs hindcast and forecast tasks qualitatively, quantitatively, and statistically.


\subsection{Turbulent jet} \label{jet}

We next consider the large-eddy simulation (LES) data of a turbulent, iso-thermal jet at Mach number of $M=0.9$, defined based on the jet velocity and the far-field speed of sound, by Br{\`e}s $\textit{et al.}$ \cite{bresetal_2018jfm}. The jet is further characterized by a Reynolds number of $\Re\approx10^6$, based on the nozzle diameter and the jet velocity. Owing to the rotational symmetry of the jet, we decompose the data into azimuthal Fourier components, $m$.
For demonstration, we take the axisymmetric component with $m=0$ as an example and interpolate it on a $950\times195$ Cartesian mesh that includes the physical domain $x,r \in [0,30] \times [0,6]$. The same dataset has been used to demonstrate the SPOD-based Galerkin model in Chu $\&$ Schmidt \cite{chu2021stochastic}, which requires the knowledge of the linearized Navier-Stokes operator and all primitive variables. In this work, we only take the pressure field as the representation of the flow state to assess the proposed data-driven model. The weighted pressure 2-norm
is used to quantify the flow energy in equation (\ref{energy}). 

	\begin{figure*}[ht!]
		\centering
        \includegraphics[trim = 0mm 0mm 0mm 0mm, clip,width=.9\textwidth]{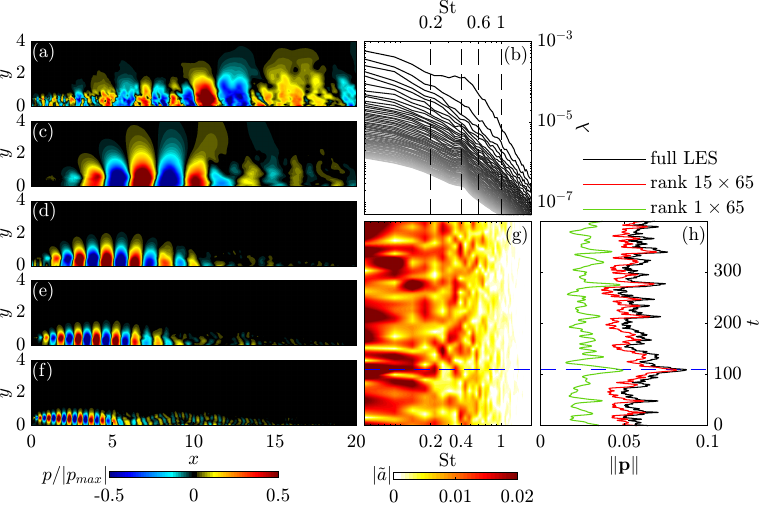}
        \caption{Overview of the turbulent jet: (a) instantaneous pressure field (marked as blue dashed lines in (g,h)); (b) SPOD eigenvalue spectra; (c-f)
        Leading SPOD modes for (c) St$=0.2$, (d) St$=0.4$, (e) St$=0.6$, and (f) St$=1$ (marked as black dashed lines in (b));
        (g) frequency-time diagram constructed from $|\tilde{a}|$; (h) time traces of the weighted pressure 2-norm.
        } \label{jet_nfft_128}
	\end{figure*}
 \vskip -2ex
 
Figure \ref{jet_nfft_128} provides a comprehensive overview of the turbulent jet data.
Here, SPOD is performed using $N=10000$ snapshots at a time step of $\Delta t =0.2$, a block size of $2N_f=128$, and an overlap of $75\%$.
The frequency is reported in terms of the Strouhal number $\mathrm{St}=\omega/(2\pi M)$, where $\omega$ is the non-dimensionalized frequency. Broadband spectrum with no tones can be observed in panel \ref{jet_nfft_128}(b), indicating that all frequencies are active in the flow, making it hard to model. The leading eigenvalues separate from the others, with maximum separation occurring at $\mathrm{St}\simeq 0.4$. This low-rank behavior has its physical origins in hydrodynamic stability. For the physical interpretation of the SPOD modes, refer to Schmidt $\textit{et al.}$ \cite{schmidt2018spectral}. While the exact flow physics are not relevant in this context, we do expect them to be relevant to reduced-order modeling. Over-proportional energy is contained in the leading modes. The first 15 modes of all $65$ non-negative frequency components contain $83.6\%$ of the energy. As shown in panel \ref{jet_nfft_128}(h), the rank $15\times65$ data, $\check{\vb*{p}}_{15\times 65}$, provides a promising reconstruction that accurately captures the dynamics of the full LES data, $\check{\vb*{p}}_{\text{full}}$. Based on this observation, it is clear that a predictive model capable of accurately representing the full data requires this number of modes.
The peak in panel \ref{jet_nfft_128}(g,h) is associated with the energetic low-frequency large-scale coherent structures in the downstream half of the computational domain, see panel \ref{jet_nfft_128}(a). From these three panels, the intermittent and stochastic nature of the turbulent jet becomes apparent, indicating that any model aiming to correctly represent this flow must be stochastic. It is pointless to seek a deterministic model that predicts the trajectory, as overfitting of the trajectory becomes unavoidable. The SLICK model is trained using $N=9000$ consecutive snapshots. The L2 regularizations are performed with $\gamma_1/\mathrm{E}\{\|\vb*{a}\|^2\}= 10^{-3}$ and $\gamma_2/\mathrm{E}\{\|\vb*{y}\|^2\}=10^{-4}$ for following our best practices.





   \begin{figure}[ht!]
		\centering
        \includegraphics[trim = 0mm 0mm 0mm 0mm, width=0.8\textwidth]{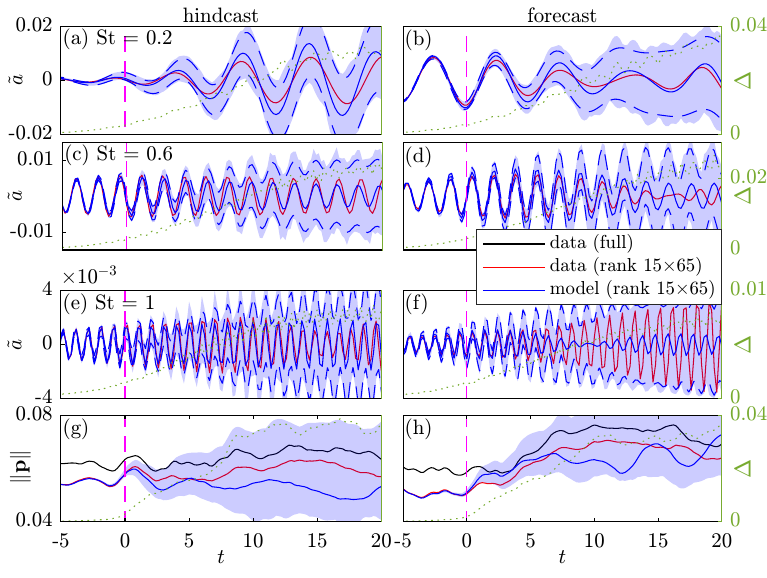}
        \caption{
Hindcast (a,c,e,g) and forecast (b,d,f,h) for the turbulent jet 
in terms of the leading $\tilde{\vb*{a}}$ at frequencies: St=0.2 (a,b), St=0.6 (c,d), and St=1 (e,f), and the integral pressure norm (g,h).
        Refer to figure \ref{SCGL_a_all} for the interpretation of symbols. Monte Carlo envelopes are generated based on 3000 model evaluations.
        } \label{jet_a_all}
	\end{figure}
 
\begin{figure}[ht!]
		\centering
        \includegraphics[trim = 0mm 12mm 0mm 0mm, width=0.85\textwidth]{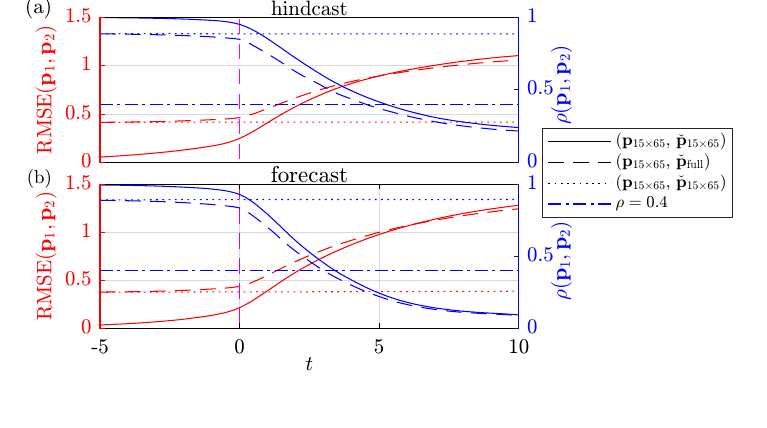}
        \caption{Normalized RMSE (left axes, red) and correlation (right axes, blue) for the hindcast (a) and forecast (b) of the jet.
        Refer to figure \ref{SCGL_error} for the interpretation of symbols. 
For the SLICK model, 500 initial conditions are each run with 50 Monte Carlo simulations. } \label{jet_error}
	\end{figure}

     Similar to figure \ref{SCGL_a_all}, we compare the rank $15\times 65$ data and a single realization of the rank $15\times 65$ model in terms of $\tilde{\vb*{a}}$ in figure \ref{jet_a_all}. For both hindcast and forecast, the specific realization of the model follows the reduced-order data for multiple periods without significant amplitude and phase error for all frequencies. The uncertainty envelope resembles the oscillatory trace of the convolutional coordinates for at least approximately ten-time units.
     Consistent with this observation, the Monte-Carlo envelope of the integral of the pressure norm starts to saturate at $t\simeq 10$. We later demonstrate that this also occurs within the context of figure \ref{jet_error}, which we thus include as the point at which the model decorrelates from the initial condition.

	\begin{figure*}[ht!]
		\centering
        \includegraphics[trim = 0mm 0mm 0mm 0mm, clip, width=.85\textwidth]{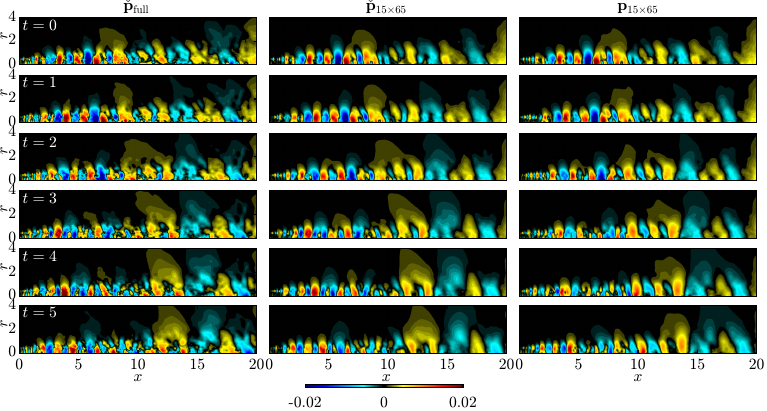}
        \caption{
        Pressure field hindcast for the turbulent jet at different leading times:   
        (left) full LES data; (middle) the rank $15\times 65$ data; (right) a random realization of SLICK.}\label{jet_snapshot_hindcast}
	\end{figure*}


 	\begin{figure*}[ht!]
		\centering
        \includegraphics[trim = 0mm 0mm 0mm 0mm, clip, width=.85\textwidth]{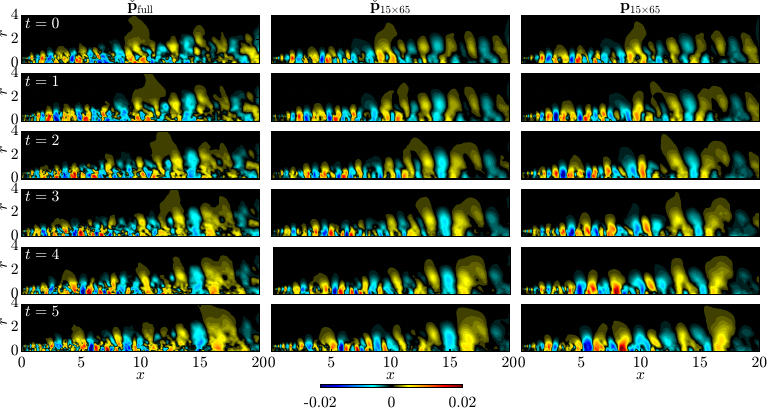}
        \caption{
        Same as figure \ref{jet_snapshot_hindcast} but for forecast.}\label{jet_snapshot_forecast}
	\end{figure*}

To examine the contributions of these convolutional coordinates to the surrogate flow field,
we compare the instantaneous pressure fields from $t=0$ to $t=5$ for the full LES data, the rank $15\times65$ data, and the model. Figure \ref{jet_snapshot_hindcast} shows the results for hindcast. It can be observed that the model closely resembles the low-rank data (and therefore the full data) up to $t\simeq2$, and even at $t=5$, it accurately captures many prominent features of the pressure field. 
Due to the convective nature of the jet, the similarity between the data and model of the faster-traveling and upstream patterns, with their significantly higher decay rate, fades more quickly. The corresponding realizations for the forecast are shown in figure \ref{jet_snapshot_forecast}.
 The model well-predicts the data up to $ t \simeq 3$, compared to $ t \simeq 5$ for hindcast.
As one would expect the challenge for real-time prediction for turbulent flows with a high Reynolds number, SLICK shows more discrepancy for later times.
However, it is worth noting that the larger flow structures toward the end of the domain are arguably well-predicted throughout the entire time interval under consideration.

 	\begin{figure*}[ht!]
		\centering
        \includegraphics[trim = 0mm 0mm 0mm 0mm, clip, width=.8\textwidth]{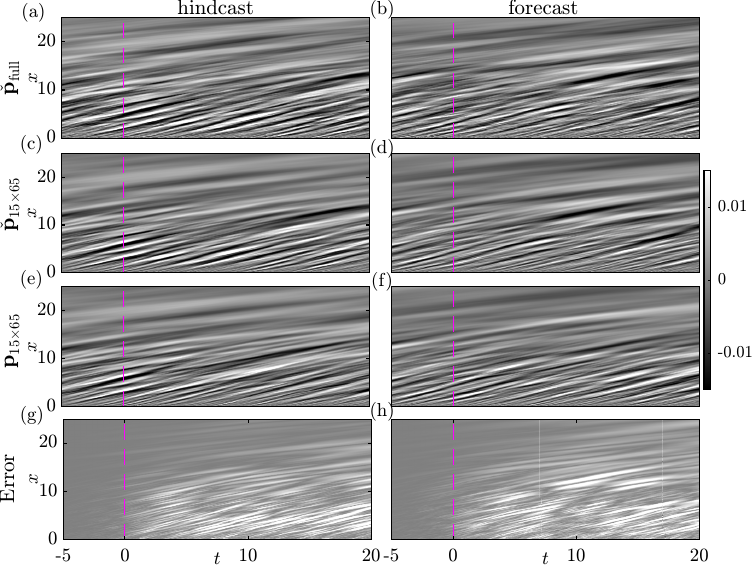}
        \caption{Comparisons between the full LES data $\check{\vb*{p}}_{\text{full}}$ (a,b), the rank $15\times65$ data $\check{\vb*{p}}_{15\times 65}$ (c,d), SLICK ${\vb*{p}}_{15\times 65}$ (e-f), and the absolute error between $\check{\vb*{p}}_{15\times 65}$ and ${\vb*{p}}_{15\times 65}$ (g-h) for hindcast (a,c,e,g) and forecast (b,d,f,h) at $r=0.5$ for different times.}\label{r=0.5}
	\end{figure*}

    To examine the spatial-temporal evolution of the flow field,
    we show the $x$-$t$ diagrams at $r=0.5$ (along the lip line) in figure \ref{r=0.5}. The convective characteristics of the flow field manifest through the diagonal pattern associated with the propagation of the wavepackets previously seen in figures \ref{jet_snapshot_hindcast} and \ref{jet_snapshot_forecast}. Visual inspection of the pressure fields suggests that SLICK accurately predicts the initial transient dynamics up to $t\simeq 5$ for the hindcast and around $t\simeq 3$ for the forecast. To quantify this observation, the error between SLICK and the low-rank data is reported in panels \ref{r=0.5} (g, h). These error plots confirm two of our previous observations. First, the overall hindcast error is somewhat lower compared to the forecast. Second, the forecast and hindcast errors are significantly lower at downstream locations ($x > 15$). It is also important to note that the strong correspondence between the model and data is not directly reflected in the error. This leads us to the conclusion that the error is mainly due to the phase discrepancy.







 Figure \ref{jet_error} shows the normalized RMSE and Pearson correlation for the hindcast and forecast of the jet.
The correlation coefficients are around 0.4 at $t \simeq 5$ for the hindcast and $t \simeq 3$ for the forecast, quantitatively confirming the corresponding prediction horizons for each case. 
This implies that SLICK for the jet would necessitate over 9000 training snapshots to achieve comparable performances for hindcast and forecast, as was the case for SCGL in figure \ref{SCGL_error}, where a large amount of data is readily available.

	\begin{figure}[ht!]
		\centering
        \includegraphics[trim = 0mm 0mm 0mm 0mm, clip,width=.4\textwidth]{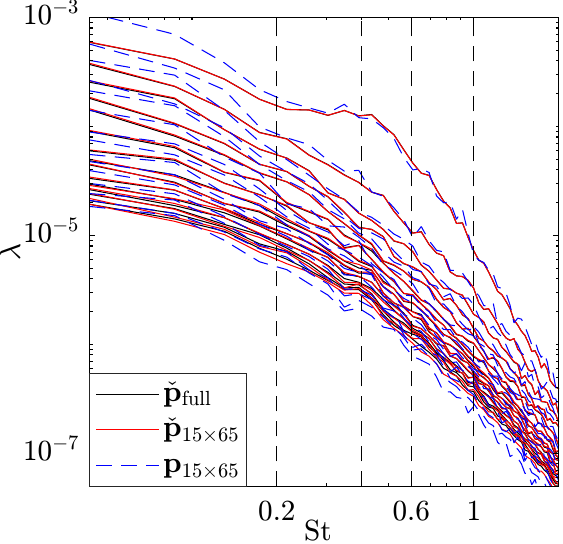}
\caption{Comparison of SPOD eigenvalue spectra of the LES data (black), the rank $15\times65$ data (red), and SLICK (blue).
} \label{spectra_jet_pressure}
	\end{figure}


   Finally, we validate that the model accurately reproduces the second-order statistics. To this end, figure \ref{spectra_jet_pressure} compares the SPOD eigenvalue spectra of the surrogate flow field produced by the model with those of both the full LES and low-rank data. 
Except for minor deviations at low frequencies, the model closely matches the eigenvalue spectra of the LES data, especially in the frequency range $0.2 \lesssim \mathrm{St} \lesssim 1.2$, where different modal and nonmodal growth mechanisms as well as acoustic resonant and nonresonant acoustic waves coexist and interact \citep{garnaud2013preferred,schmidt2017wavepackets,schmidt2018spectral,tissot2017sensitivity,towne2017acoustic}.
This suggests that the surrogate flow field preserves the underlying physical mechanisms.

\subsection{Open cavity flow} \label{cavity}

	\begin{figure*}[ht!]
		\centering
        \includegraphics[trim = 0mm 0mm 00mm 0mm, clip,width=.8\textwidth]{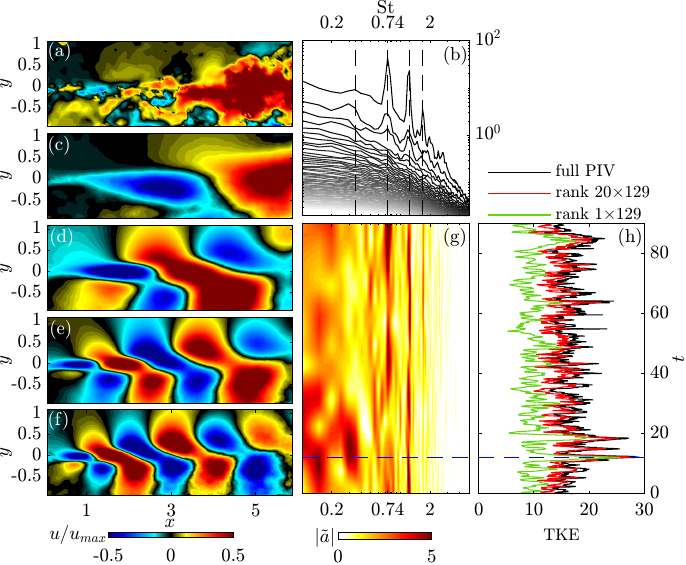}
        \caption{Overview of the open cavity flow: (a) instantaneous streamwise velocity (marked as blue dashed lines in (g,h)); (b) SPOD eigenvalue spectra; (c-f) leading SPOD modes at the first $4$ Rossiter frequencies (black dashed lines in (b)): (c) St$=0.34$; (d) St$=0.74$; (e) St$=1.23$; (f) St$=1.67$;
        (g) frequency-time diagram; (h) time traces of TKE.
        } \label{cavity_nfft_256}
	\end{figure*}

\begin{figure}[ht!]
		\centering
        \includegraphics[trim = 0mm 0mm 0mm 0mm, clip,width=.425\textwidth]{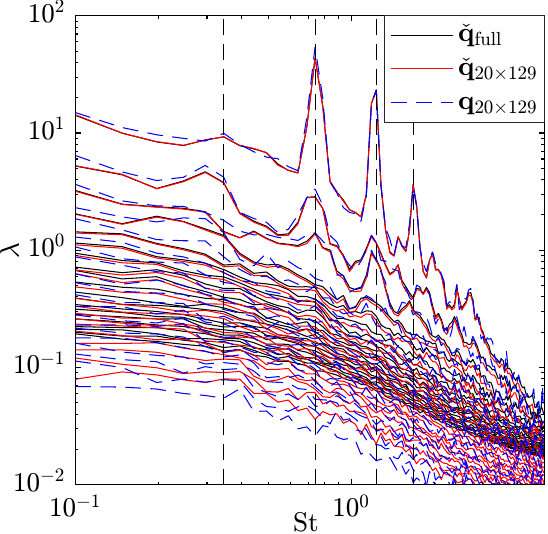}
\caption{Comparison of SPOD eigenvalue spectra of the full PIV data (black), the rank $20\times129$ data (red), and SLICK (blue dashed).
} \label{spectral_cavity}
	\end{figure}

 
Our last example is the experimental data of a high-Reynolds number flow over an open cavity obtained from time-resolved particle image velocimetry (TR-PIV)
by Zhang \textit{et al.}\cite{zhang2017identification}.
This case is representative of real-world data of complex, mixed-broadband-tonal turbulent flows in the presence of measurement noise.
The center plane of the open cavity has a length-to-depth ratio of $L/D=6$ and a width-to-depth ratio of $W/D=3.85$. The Reynolds number based on the cavity depth is $\Re=\rho U_{\infty}D/\mu \approx 3.3\times 10^5$, and the Mach number is $M=U_{\infty}/c_{\infty}=0.6$, where  $c_{\infty}$ is the far-field speed of sound.
The sampling rate is 16 kHz. 
For further details, refer to Zhang \textit{et al.} \cite{zhang2017identification,zhang2020spectral}. We non-dimensionalize the spatial coordinates, velocity, and time by $D$, $U_{\infty}$, and $L/U_{\infty}$, respectively. The Strouhal number is defined as $\mathrm{St} =f L /U_{\infty}$. The compound velocity field, $\vb*{q} =\left[u,\,v\right]^T $, is chosen as the flow state. The flow energy in equation (\ref{energy}) is hence the turbulent kinetic energy, $\mathrm{TKE}=\frac{1}{2}\int_{\Omega}\left(u'^2+v'^2\right) \dd \vb*{x} $.
 A comprehensive overview of the open cavity flow is provided in figure \ref{cavity_nfft_256}.
  SPOD is performed using $N=16000$ snapshots with a block size of $2N_f=256$ and an overlap of $75\%$.
These parameters are found to provide a good balance between the salience of physics, that is, the Rossiter frequencies, and the computational cost of the model. The resulting SPOD eigenvalue spectra in panel \ref{spectral_cavity}(b) match well with those reported by  \cite{zhang2020spectral,schmidt2022spectral}, particularly the dominant second and third Rossiter tones.  
As shown in the frequency-time diagram in panel \ref{spectral_cavity}(g), the 3rd Rossiter tone is the most persistent over the entire time horizon, while the rest appear somewhat intermittent.
The SLICK model is constructed using the first 20 modes at all $129$ non-negative frequencies, containing $76.6\%$ of the SPOD energy.
The corresponding rank $20\times129$ data, also shown in panel \ref{cavity_nfft_256}(h),
accurately captures the dynamics of the full data. The training set comprises $N=14400$ snapshots.
The same ridge parameters in \S\ref{jet}, $\gamma_1/\mathrm{E}\{\|\vb*{a}\|^2\}= 10^{-3}$ and $\gamma_2/\mathrm{E}\{\|\vb*{y}\|^2\}=10^{-4}$, are used for L2 regularization.

    \begin{figure}[ht!]
		\centering
        \includegraphics[trim = 0mm 0mm 0mm 0mm, width=0.75\textwidth]{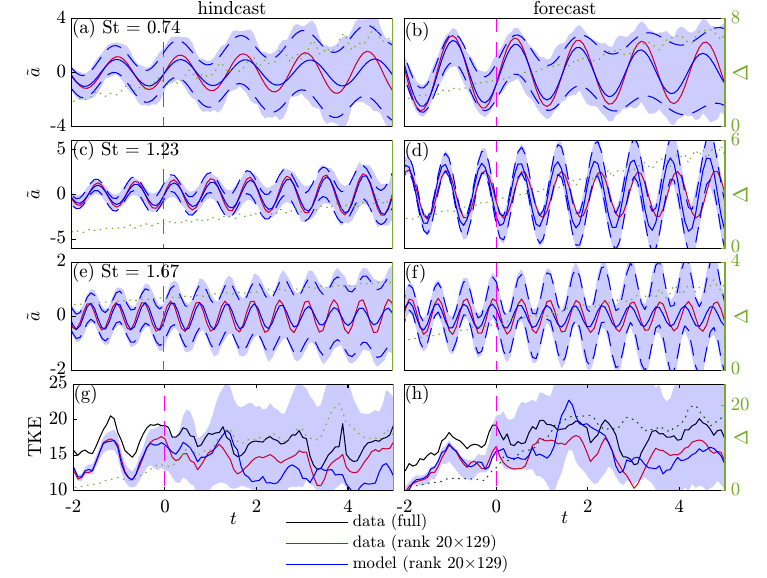}
        \caption{
        Hindcast (a,c,e,g) and forecast (b,d,f,h) for the open cavity flow
in terms of the leading $\tilde{\vb*{a}}$ at three peak Rossiter frequencies: St=0.74 (a,b), St=1.23 (c,d), and St=1.67 (e,f), and the total TKE (g,h). Refer to figure \ref{SCGL_a_all} for the interpretation of symbols. Monte Carlo envelopes are generated based on 3000 model evaluations.
        } \label{cavity_a_all}
	\end{figure}

 
Figure \ref{cavity_a_all} compares the trajectories of the rank $20\times 129$ model to the data for both the hindcast and forecast.
In both cases, the shown realization demonstrates favorable agreements with the low-rank data.
Compared to the SCGL and jet cases (see figures \ref{SCGL_a_all} and \ref{jet_a_all}), the model exhibits more uncertainty in the total energy at $t=0$.
This is attributed to the presence of measurement noise, gaps, and artifacts within the PIV measurements, leading to increased uncertainty in inferring the system dynamics from the data.

	\begin{figure}
		\centering
        \includegraphics[trim = 0mm 0mm 0mm 0mm, clip, width=1\textwidth]{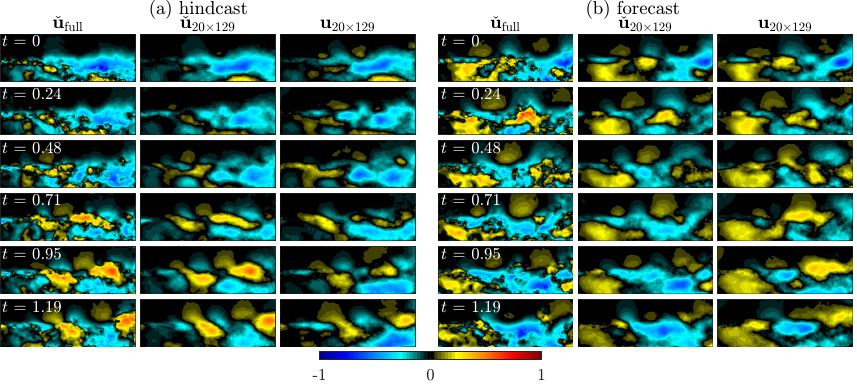}
        \caption{
        Streamwise velocity field hindcast (a) and forecast (b) for the open cavity flow within the domain $x,\,y \in [0,\,6]\,\times\,[-1,\,1]$ at different leading times.   
        }\label{cavity_snapshot_all}
	\end{figure}

\begin{figure}[ht!]
		\centering
        \includegraphics[trim = 5mm 13mm 10mm 0mm, width=0.75\textwidth]{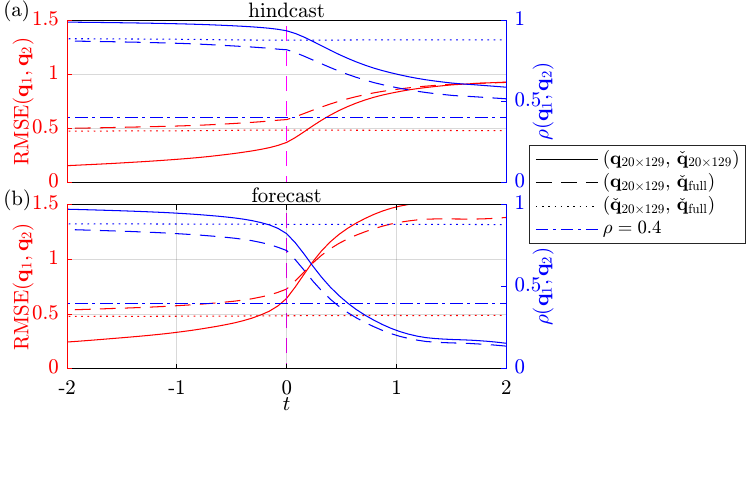}
        \caption{Normalized RMSE (left axes, red) and Pearson correlation (right axes, blue) for hindcast (a) and forecast (b) of the open cavity flow. 
               Refer to figure \ref{SCGL_error} for the interpretation of symbols. 
For the SLICK model, 300 initial conditions are each run with 50 Monte Carlo simulations. } \label{cavity_error}
	\end{figure}

Figure \ref{cavity_snapshot_all} shows the instantaneous velocity fields from $t=0$ to $t=1.19$ for $\check{\vb*{u}}_{\text{full}}$, $\check{\vb*{u}}_{20\times 129}$, and ${\vb*{u}}_{20\times 129}$ for both the hindcast and forecast.
For the hindcast, the model output closely resembles the data up to about half of a flow-through time, $t\simeq0.48$.
However, even at $t=1.19$, many prominent features of the streamwise velocity field are still present.
As expected, the similarity between the model and the low-rank data diminishes more rapidly for the forecast, with a good qualitative correspondence observed only for $t\leq0.24$. These subjective observations will be later quantified in terms of the Pearson correlation coefficient, as discussed in the context of figure \ref{cavity_error}.

 	\begin{figure}[ht]
		\centering
        \includegraphics[trim = 0mm 0mm 0mm 0mm, clip, width=.8\textwidth]{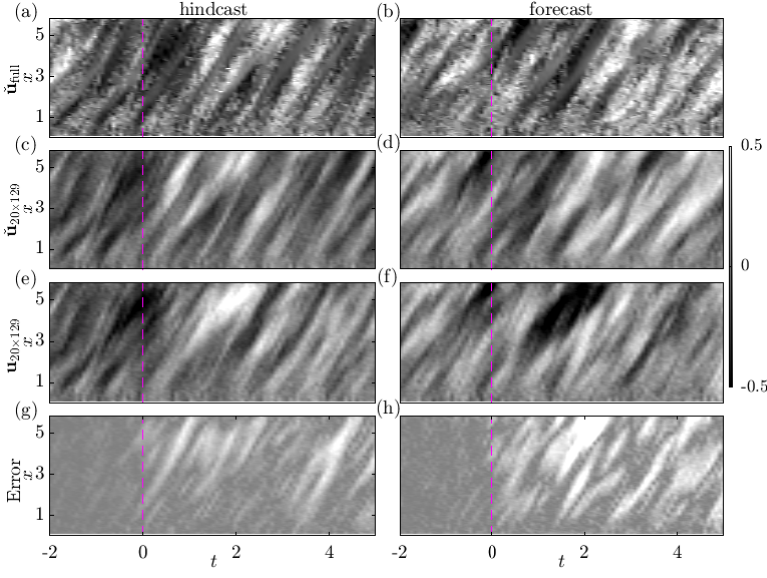}
        \caption{Comparisons between the full PIV data $\check{\vb*{u}}_{\text{full}}$ (a,b), the rank $20\times129$ data $\check{\vb*{u}}_{20\times129}$ (c,d), SLICK ${\vb*{u}}_{20\times129}$ (e-f) and the absolute error between $\check{\vb*{u}}_{20\times 129}$ and ${\vb*{u}}_{20\times 129}$ (g-h) for the hindcast (a,c,e,g) and the forecast (b,d,f,h) at $y=0$ for different times.  }\label{y=0}
	\end{figure}

 Similar to figure \ref{r=0.5}, we show the $x$-$t$ diagrams at $y=0$  for both the hindcast and forecast in figure \ref{y=0}. The model output closely resembles the general convective behavior of the flow field.
Consistent with figure \ref{cavity_snapshot_all}, the model for the hindcast has a longer prediction horizon than the forecast.
The hindcast closely mimics the details of the flow field until 
$t=2$, with deviations observed somewhat earlier in the forecast.

  Figure \ref{cavity_error} quantifies the error between the model output and the data.
Compared to figures \ref{SCGL_error} and \ref{jet_error}, the overall model performance declines
due to measurement noise and gaps within the data. 
These issues are intrinsic to PIV measurements, rendering the prediction of PIV data a particularly challenging endeavor.
Despite this, our model yields fair predictions ($\rho\gtrsim0.4$) for hindcasting with at least 2 flow-through times and forecasting at $t\simeq0.5$.

 Figure \ref{spectral_cavity} compares the SPOD eigenvalue spectra of SLICK with those of both $\check{\vb*{q}}_{20\times 129}$ and $\check{\vb*{q}}_{\text{full}}$. 
Favorable comparisons are observed, particularly at the peak Rossiter frequencies, showing that the second-order flow statistics are accurately captured.
The above results demonstrate SLICK's capability for both short-term prediction and long-term statistical reproduction.

\section{Summary and discussion}	\label{conclusion}

In this work, SLICK, a stochastic data-driven reduced-order model, is developed to accurately predict turbulent flows, with a focus on capturing both short-time transient dynamics and long-time statistics. SLICK describes the motion of large-scale coherent structures by propagating inflated SPOD convolutional coordinates in time using a finite-dimensional approximation of the Koopman operator.
The residual error inherent in approximating turbulent dynamics is modeled as a stochastic source, wherein a data-informed dewhitening filter is employed to maintain second-order flow statistics. SLICK has been demonstrated using one numerical dataset and two fundamental, large, research-grade, state-of-the-art fluid mechanics datasets.
Both qualitative and quantitative assessments, coupled with uncertainty quantification, affirm the capability of SLICK to accurately reproduce the dynamics and statistics of turbulent flows without the need for governing equations.

One major contribution of this work is the incorporation of 
spectral proper orthogonal modes and the time-delay Koopman framework.
We propose interpreting SPOD expansion coefficients as convolutional coordinates and the latter, in turn, as time-delay observables.
This re-interpretation consequently results in an algorithm for computing these coefficients with significantly lower computational cost compared to the standard algorithm \citep{nekkanti2021frequency}.
This perspective can alternatively be elucidated through the inherent relationship between the Hankel singular vectors and SPOD modes \citep{frame2023space}.
Consequently, SLICK also establishes connections to the HAVOK model \citep{brunton2017chaos}, wherein non-Gaussian intermittent forcing is employed to drive the dynamical system.
Instead of considering intermittent forcing, the inflation of state vectors in SLICK arises from the most fundamental perspective of Koopman theory: incorporating nonlinear observables alongside linear ones is necessary for comprehensive nonlinear system characterization.
This viewpoint also facilitates the interpretation of established MLR models \cite{kondrashov2005hierarchy,kravtsov2005multilevel} within the Koopman framework, where each level of the linear model inflates the state vector
based on the residual nonlinear terms from the previous state.


The computational cost of SLICK mainly depends on solving a system of stochastic ODE with dimensions of $2 M \times (N_f + 1)$.
This enables the use of Monte Carlo simulations for uncertainty quantification.
One prospective future development is the incorporation of SLICK with optimal stochastic control to improve forecasts of turbulent flows.
In particular, the use of linear form and Gaussian white noise in SLICK
makes the Kalman filter well-suited for this purpose.

\section*{Data access}
{Code available at \url{https://github.com/SLICK-model/SLICK_matlab}.}

\section*{CRediT authorship contribution statement}
{T.C.: conceptualization, data curation, formal analysis, investigation, methodology, software, validation, visualization, writing--original draft, writing–-review $\&$ editing; O.T.S.: conceptualization, investigation, methodology, project administration, resources, supervision, writing--original draft, writing--review $\&$ editing.}

\section*{Declaration of competing interest}
{We declare we have no competing interests.}

\section*{Acknowledgments}
{We received no funding for this study.}
  



 \bibliographystyle{RS}

\bibliography{references}

\newpage \clearpage

\begin{appendix}

\begin{center}%
 {\Large \mathversion{bold}Supplementary Material}%
\end{center}

\setcounter{page}{1}
\setcounter{section}{0}
\setcounter{figure}{0}
\renewcommand{\theequation}{S.\arabic{section}.\arabic{equation}}

\textit{
Please note that this supplementary file should be read in conjunction with the main manuscript.
}

\section{A review of modal decomposition/operator-based techniques}

    	\begin{figure*}[ht!]
		\centering
        \includegraphics[trim = 10mm 5mm 5mm 5mm, width=1\textwidth]{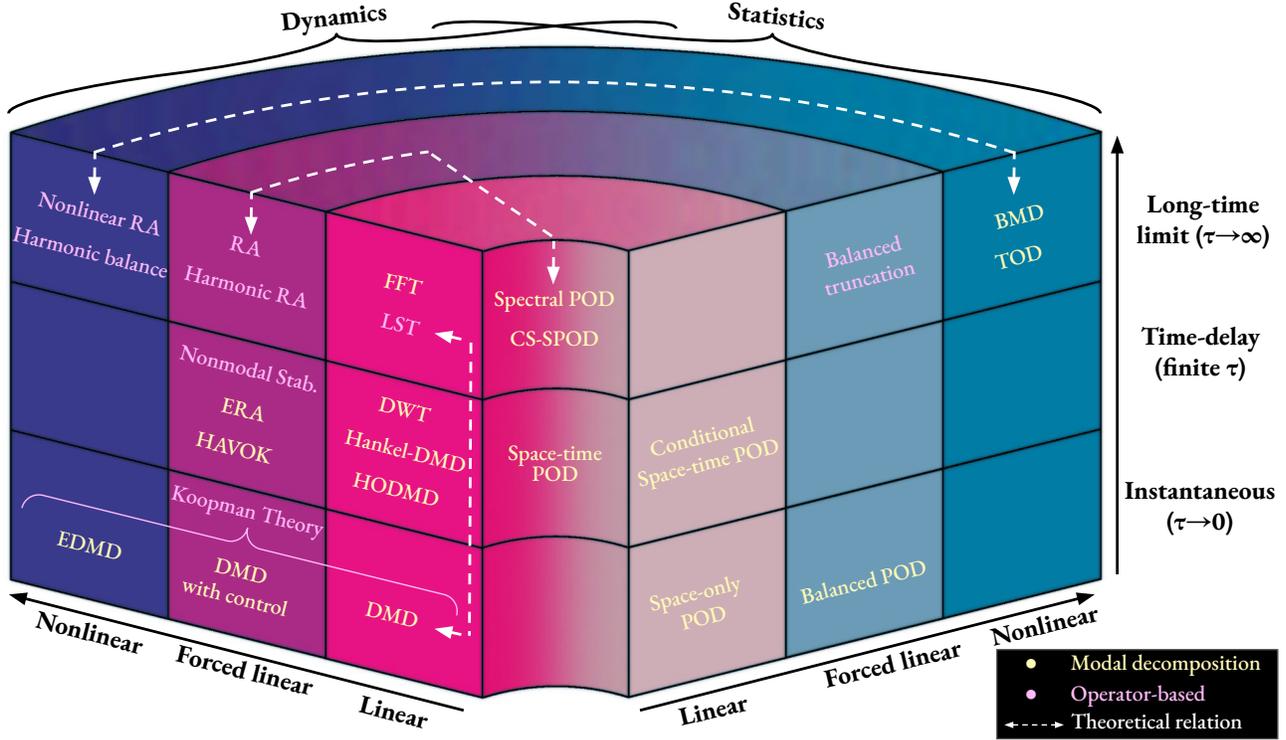} 
        \caption{Classification of 
        data-driven and operator-based methods for basis identification methods for model order reduction, categorized according to three modeling perspectives: 
        (i) analysis type (statistical, dynamical, or hybrid), (ii) state vector structure (linear $\to$ forced linear $\to$ nonlinear), and (iii) time-delay horizon (instantaneous $\to$ finite-time $\to$ long-time limit).
        }  
	\end{figure*}



This paragraph provides a succinct summary of established basis identification methods for model order reduction, elucidating their interrelation with spectral proper orthogonal decomposition (SPOD)~\citep{towne2018spectral_app, schmidt2018spectral_app} and Koopman operator theory.
Incorporating the time-delay characteristic inherent in SPOD convolutional coordinates with Koopman theory culminates in the development of the proposed stochastic
low-dimensional inflated convolutional Koopman (SLICK) model.
Operator-based approaches derive modal bases from discretized governing operators, whereas modal decomposition techniques extract them from simulation or experimental data snapshots.
We categorize these techniques from three perspectives. 
The first perspective distinguishes between statistical and dynamic modeling approaches based on whether flow dynamics are encoded within the modes.
The most popular statistical framework for model order reduction is the classical proper orthogonal decomposition (POD) \citep{lumley1967structure_app,lumley2007stochastic_app}.
In statistically based models, temporal information is separated from spatial modes and is represented as time-dependent expansion coefficients.
Dynamic mode decomposition (DMD) \citep{schmid2010dynamic_app}, on the other hand, postulates that the flow state evolves linearly over time, capturing the temporal behavior of each mode as a complex frequency that encodes both oscillation and growth or decay rates.
Both methods use instantaneous flow fields, or snapshots, as observables to represent nonlinear flows, storing them as linear state vectors.
As all real-world technical and environmental flows are inherently nonlinear, various approaches have been developed to incorporate state-dependent but time-invariant forcing, enhancing the representation of nonlinear effects in the evolution of state vectors.
This leads to the second perspective in modeling general fluid flows, namely, the representation of a nonlinear flow field.
Examples include balanced POD \citep{rowley2005model_app} for POD, and DMD with control \citep{brunton2015closed_app,brunton2016koopman_app} for DMD.
Based on Koopman theory \citep{koopman1931hamiltonian_app}, extended DMD (EDMD) \citep{williams2015data_app} directly includes nonlinear observables of the flow field into the state vector.
While all of the above approaches use instantaneous flow states, time-delay embedding of the flow states provides an alternative way of incorporating spatiotemporal nonlinearity, thereby presenting the last perspective for categorizing these methodologies.
With a finite time-delay horizon, Hankel-DMD \citep{arbabi2017ergodic_app} and higher-order DMD (HODMD) \citep{le2017higher_app} extend the conventional DMD method by employing a Hankelized approach.
The integration of time-delay embedding and time-invariant forcing for model order reduction can be traced back to the eigensystem realization algorithm (ERA)~\citep{juang1985eigensystem_app}.
Nonmodal stability theory~\citep{schmid2007nonmodal_app} provides a theoretical methodology for modeling transient growth, while the Hankel alternative view of Koopman (HAVOK) analysis \citep{brunton2017chaos_app} incorporates an exogenous forcing term within the time-delay DMD framework.
As a variant of POD, conditional space-time POD \citep{schmidt2019conditional_app} has been proposed to extract the dominant spatiotemporal structures within a finite time window.
Building upon this, space-time POD \citep{frame2023space_app} provides both statistical and dynamical descriptions of the flow field and has been shown to converge to SPOD in the long-time limit.
This property makes SPOD a well-suited choice for model order reduction.
In this limit, the time-delay embedding converges to the Fourier transform, and 
Fourier series can be used instead of localized bases like wavelets~\citep{farge1992wavelet} for statistically stationary flows.
Numerous hydrodynamic stability theories have been well-established
to obtain modal bases from discretized linearized operators. 
Classical linear stability analysis (LST) assumes the exponential modal growth of small perturbations, which aligns with the underlying principles of DMD.
Rooted in early studies of transient growth~\citep{trefethen1993hydrodynamic,schmid2002stability}, resolvent analysis (RA) \citep{jovanovic2005componentwise_app, hwang2010amplification_app, mckeon2010critical_app} has evolved into a linear tool for predicting instabilities of turbulent flows. 
It models nonlinear interactions, along with the background turbulence, as external forcing to the otherwise linear dynamics.
Theoretical correspondence has been established between RA and SPOD \citep{towne2018spectral_app}, as well as their variants for periodic flows, namely Harmonic RA~\citep{padovan2020analysis_app} and cyclostationary SPOD (CS-SPOD)~\citep{heidt2023spectral_app}. 
Theoretical approaches include Harmonic Balance~\citep{hall2002computation} and Nonlinear RA~\citep{rigas2021nonlinear_app}, while bispectral mode decomposition (BMD)~\citep{schmidt2020bispectral_app} and triadic orthogonal decomposition (TOD)~\citep{yeung2024revealing_app} offer data-driven ways to identify triadic interactions.



\section{Deriviation of SPOD expansion coefficients as optimal convolutional coordinates}

 We decompose the fluctuating state into its temporal discrete Fourier modes, $\hat{\left(\cdot\right)}$, as
\begin{align}\label{q_DFT_App}
    \vb*{q}'(t) = \frac{1}{2N_f\Delta t}\sum_{l=1}^{2N_f} \hat{\vb*{q}}(\omega_l) \mathrm{e}^{\mathrm{i} \omega_l t}
\end{align}
on the time interval $(-N_f \Delta t,N_f \Delta t]$,
where $\omega_l =  l \pi /(N_f \Delta t)$ is the angular frequency. 
Inserting $  \phi_j(\vb*{q}) = \left<\vb*{\psi}_j,\vb*{q}'\right>_{x}$ into equation (\ref{q_DFT_App}) yields the $j$th observable
\begin{align} \label{phi_j_app}
    \phi_j(\vb*{q}) = \frac{1}{2N_f\Delta t}\sum_{l=1}^{2N_f}\left<\vb*{\psi}_j,\hat{\vb*{q}} (\omega_l)\right>_x \mathrm{e}^{\mathrm{i} \omega_l t}.
\end{align}
Recall the discrete convolution of the $j$th observable on the time interval $(-N_f \Delta t,N_f \Delta t]$,
\begin{align} \label{conv_observable_App}
    {a}_j[i]=  \sum_{h=1-N_f}^{N_f} c_{h,j} \phi_j[i+h],
\end{align}
where the convolutional weights are defined as
\begin{align}\label{conv_w_App}
    c_{h,j} \equiv \mathrm{e}^{-\mathrm{i}\tilde{\omega}_j h\Delta t}w[h\Delta t]\Delta t.
\end{align}
Combining equations (\ref{q_DFT_App}-\ref{conv_w_App}) leads to
\begin{align}
    {a}_j[i] 
    & = \sum_{h=1-N_f}^{N_f} c_{h,j} \left( \frac{1}{2N_f\Delta t}\sum_{l=1}^{2N_f}\left<\vb*{\psi}_j,\hat{\vb*{q}} (\omega_l)\right>_x \mathrm{e}^{\mathrm{i} \omega_l (i+h)\Delta t} \right)  \nonumber \\
    &= \frac{1}{2N_f\Delta t} \sum_{l=1}^{2N_f} \left<\vb*{\psi}_j,\hat{\vb*{q}}(\omega_l)\right>_x   \left(\sum_{h=1-N_f}^{N_f} c_{h,j} \, \mathrm{e}^{\mathrm{i} \omega_l (i+h)\Delta t}\right)   \nonumber \\
    & =  \frac{1}{2N_f\Delta t} \sum_{l=1}^{2N_f} \left<\vb*{\psi}_j,\hat{\vb*{q}}(\omega_l)\right>_x   \left(\Delta t \, \mathrm{e}^{\mathrm{i} \omega_l (i\Delta t)}\sum_{h=1-N_f}^{N_f} \mathrm{e}^{\mathrm{i} (\omega_l-\tilde{\omega}_j) h\Delta t}w[h\Delta t]\right) \nonumber\\
    & = \sum_{l=1}^{2 N_f} \left<\vb*{\psi}_j,\hat{\vb*{q}}(\omega_l)\right>_x \mathrm{e}^{\mathrm{i} \omega_l (i \Delta t)} \delta(\omega_l-\tilde{\omega}_j) \nonumber\\   
    &=   \left<\vb*{\psi}_{j},\hat{\vb*{q}}(\tilde{\omega}_j)\right>_x \mathrm{e}^{\mathrm{i} \tilde{\omega}_j (i \Delta t)},
\end{align}
where $\delta(\cdot)$ represents the Dirac delta function. The magnitude of $a_j$ only depends on the inner product of the $j$th basis vector, $\vb*{\psi}_{j}$, and the Fourier mode associated with frequency $\tilde{\omega}_j$, $\hat{\vb*{q}}(\tilde{\omega}_j)$.
With modes that have unit energy, $ {\|\vb*{\psi}_{j}\|_x=1}$, we seek the convolutional coordinates that optimally represent the flow field at each given frequency, $\tilde{\omega}_j$. This objective is formalized by maximizing
the quantity
\begin{align}\label{Rayleigh_quotient_App}
    \lambda_j = 
    \mathrm{E}\{ a_ja_j^*\} &= 
    \mathrm{E}\{ \left( \left<\vb*{\psi}_{j},\hat{\vb*{q}}(\tilde{\omega}_j)\right>_x  \mathrm{e}^{\mathrm{i} \tilde{\omega}_j (i \Delta t)} \right)  \left( \left<\vb*{\psi}_{j},\hat{\vb*{q}}(\tilde{\omega}_j)\right>_x  \mathrm{e}^{\mathrm{i} \tilde{\omega}_j (i \Delta t)} \right)^* \} \nonumber \\
    &= 
    \mathrm{E}\{\left<\vb*{\psi}_{j},\hat{\vb*{q}}(\tilde{\omega}_j)\right>_x\left<\hat{\vb*{q}}(\tilde{\omega}_j),\vb*{\psi}_{j}\right>_x\}  \nonumber \\
    & = \mathrm{E}\{\left<\vb*{\psi}_{j},\hat{\vb*{q}}(\tilde{\omega}_j)\hat{\vb*{q}}(\tilde{\omega}_j)^* \vb*{W}\vb*{\psi}_{j} \right>_x\}  \nonumber \\
    &= \left< \vb*{\psi}_{j}, \vb*{S}_j \vb*{W} \vb*{\psi}_{j}   \right>_x, 
\end{align}
 where $\mathrm{E}\{\cdot\}$ denotes the expected value over a large number of realizations, and $\vb*{S}_j=\mathrm{E}\{ \hat{\vb*{q}}(\tilde{\omega}_j) \hat{\vb*{q}}(\tilde{\omega}_j)^* \}$ is the cross-spectral density matrix.
The maximum values of the generalized Rayleigh quotient (\ref{Rayleigh_quotient_App}) can be found from the eigenvalue problem
 \begin{align}
      \vb*{S}_j \vb*{W} \vb*{\psi}_{j}^{(\alpha)}  = \lambda_j^{(\alpha)} \vb*{\psi}_{j}^{(\alpha)},
 \end{align}
 whose eigenvectors, $\vb*{\psi}_{j}^{(\alpha)}$, are known as SPOD modes. 
This proves that the optimal
convolutional coordinates for this time-delay embedding are provided by SPOD.

\section{Flow field reconstruction}

For a given time index $i$ and a given time lag $h\in[1-N_f,N_f]$, the flow field can be reconstructed as
\begin{align}
    \vb*{q}'[i;h] &\approx  \frac{1}{2N_f\Delta t w(h\Delta t)} 
    \sum_{j=1}^{2MN_f} \left(a_j[i+h]\vb*{\psi}_{j}  \right)  \mathrm{e}^{\mathrm{i} \tilde{\omega}_j  (N_f-h)\Delta t }.
\end{align}
As the window-weighted average leads to better reconstruction \cite{nekkanti2021frequency_app}, we consider the time window $\left[h_-,\,h_+\right]\subset[1-N_f,N_f]$ such that
\begin{align} \label{q_reconst_weighted}
    \vb*{q}'[i]\approx \mathrm{E}_{h\in [h_-,h_+]} \{\vb*{q}'[i;h]\} &=  \frac{1}{2N_f\Delta t \sum_{h=h_{-}}^{h_+} w(h\Delta t)} \sum_{h=h_{-}}^{h_+} \sum_{j=1}^{2MN_f} \left(a_j[i+h]\vb*{\psi}_{j}  \right)  \mathrm{e}^{\mathrm{i} \tilde{\omega}_j  (N_f-h)\Delta t } \nonumber\\
    & = \sum_{j=1}^{2MN_f} \underbrace{\left(\frac{\sum_{h=h_{-}}^{h_+}  \left(a_j[i+h] \right)  \mathrm{e}^{\mathrm{i} \tilde{\omega}_j  (N_f-h)\Delta t }}{2N_f\Delta t \sum_{h=h_{-}}^{h_+} w(h\Delta t)}\right)}_{\tilde{a}_j[i]}\vb*{\psi}_{j},
\end{align}
which can be written in a compact form as 
\begin{align}
  \vb*{q}'[i] \approx \vb*{V}\tilde{\vb*{a}}[i].
\end{align}
Here, $\tilde{a}_j$ constitutes the $j$th component of $\tilde{\vb*{a}}$. 
Given the convolutional coordinates $a_j$, equation (\ref{q_reconst_weighted}) offers a particularly economical way for the reconstruction of flow data, as both the weighting and Fourier inversion are performed on the coordinates, not on the flow data. In practice, the weighted convolutional coordinates, $\tilde{\vb*{a}}$, are weighted through the first half of the block in equation (\ref{q_reconst_weighted}), i.e., $h_- = 1-N_f$ and $h_+ =0$. 
This arrangement confirms the causality of the model. 

\newpage \clearpage
\section{Algorithms: the stochastic low-dimensional inflated convolutional Koopman (SLICK) model} \label{algorithm}

The processing of the SLICK model is driven by three consecutive algorithms: model inference, model integration, and flow-field reconstruction, as shown below.
Matlab codes are available at \url{https://github.com/SLICK-model/SLICK_matlab}.

\begin{algorithm}
\caption{Model inference}
  \hspace*{\algorithmicindent} \textbf{Input:} Fluctuating data matrix $\vb*{Q}'=\left[
\begin{matrix}
  \vert & \vert &  &\vert \\
  \vb*{q}[1]-\overline{\vb*{q}}& \vb*{q}[2]-\overline{\vb*{q}} &\cdots & \vb*{q}[N_\text{train}]-\overline{\vb*{q}}\\
    \vert & \vert &  &\vert
\end{matrix}\right]$, time step $\Delta t$, block size $2N_f$, weight matrix $\vb*{W}$ (default as $\vb*{I}$), window function $w(\tau)$ (default as hamming window), model rank $M$, and regression parameters $\gamma_1$ and $\gamma_2$. \\
   \hspace*{\algorithmicindent} \textbf{Output:} Matrices of the SLICK model, $\vb*{T}$, $\vb*{G}$, and $\vb*{Y}$.
        \begin{algorithmic}[1]
\State  Compute SPOD of $\vb*{Q}'$ and store the first $M$ SPOD modes in the column matrix $\vb*{V}$.
 \For {$i=1:N_\text{train}-2N_f+1$} 
 \State 
 Determine the $i$th observable block as $ \vb*{\Phi}[i]  
=  \vb*{V}^* \vb*{W} \left[
\begin{matrix}
  \vert &  &\vert \\
  \vb*{q}'[i] & \cdots  &\vb*{q}'[i+2N_f-1]\\
  \vert & &\vert 
  \end{matrix} \right].$
\State Calculate the DFT using a windowed fast Fourier transform $ \hat{\vb*{\Phi}}[i]= \text{FFT}\left(w\circ \left(\vb*{\Phi}[i]\right)\right)$.
\State
 Determine the convolutional coordinates vector 
 as
$
    \vb*{a}[i+N_f-1] = \mathrm{diag}\left(\hat{\vb*{\Phi}}[i]\right).
$
\EndFor
\State Assemble the convolutional coordinates matrix $
    \vb*{A} = \left[
\begin{matrix}
  \vert &  &\vert \\
  \vb*{a}[N_f] & \cdots  &\vb*{a}[N_\text{train}-N_f]\\
  \vert & &\vert 
  \end{matrix} \right].
$
\State Determine the matrix
$
    \vb*{K} =
    (\vb*{A}_2-\vb*{A}_1) \vb*{A}_1^*\left(\vb*{A}_1 \vb*{A}_1^*+\gamma_1\vb*{I}\right)^{-1}/{\Delta t},
$
where the subscripts, $\left(\cdot\right)_1$ and $\left(\cdot\right)_2$, represent the submatrices by excluding the last and the first columns, respectively.
\State Store the data matrix of the forcing coefficients 
$    
             \vb*{B} = (\vb*{A}_2-\vb*{A}_1)/\Delta t- \vb*{K}\vb*{A}_{1},
$    
\State Assemble the inflated-state matrix $\vb*{Y}=\mqty[\vb*{A}_1^\top & \vb*{B}^\top]^\top$. 
\State Calculate the matrix 
$   
            \vb*{M} = (\vb*{B}_2-\vb*{B}_1) \vb*{Y}_1^*\left(\vb*{Y}_1 \vb*{Y}_1^*+\gamma_2\vb*{I}\right)^{-1}/{\Delta t}.
$  
           \item Assemble the matrices to obtain  
$
            \vb*{T} =  \vb*{I}+   \left[
\begin{array}{c r}
    \begin{matrix} \vb*{K}_{\vb*{a}} & \vb*{I} \end{matrix} \\
     \mathrel{\vcenter{\hbox{\rule{0.3cm}{0.5pt}}}}\vb*{M} \mathrel{\vcenter{\hbox{\rule{0.3cm}{0.5pt}}}}
\end{array}
\right] \Delta t.
$
\State
   Compute the matrix
$
  \vb*{H}= \mqty[ \vb*{0} & 
     \, \vb*{1}] \left(\vb*{Y}_2\vb*{Y}_2^* -\vb*{T}\vb*{Y}_1\vb*{Y}_1^*\vb*{T}^*\right) \mqty[ \vb*{0} & 
     \, \vb*{1}]^\top /{[(N_{\text{train}}-N_f-2)\Delta t]}.
$
\State Calculate the eigenvalue decomposition
$
    \vb*{H} = \vb*{V}_H \vb*{D}_H \vb*{V}_H^*.
$
\State Set the negative eigenvalues in $\vb*{D}_H$ to $0$ and rescale the rest positive eigenvalues such that $\mathrm{tr}(\vb*{D}_H)=\mathrm{tr}(\tilde{\vb*{D}}_H)$. 
\State
Determine the dewhitening filter matrix
$
            \vb*{G} = \vb*{V}_H \tilde{\vb*{D}}_H^{1/2}.
$
\end{algorithmic}
\end{algorithm}

\begin{algorithm}[ht]
\caption{Model integration}
  \hspace*{\algorithmicindent} \textbf{Input:} Initial condition $\vb*{y}[0]$, time step $\Delta t$, integration window size $N_t$, matrices of the SLICK model, $\vb*{T}$ and $\vb*{G}$. \\
   \hspace*{\algorithmicindent} \textbf{Output:} Inflated-state matrix $\vb*{Y}$.
\begin{algorithmic}[1]
    \For {$j = 1:N_t$}
    \State Draw a random seed.
    \State Construct the zero mean, unit variance Gaussian white noise vector using the pseudorandom number generator $\vb*{w}=\mathrm{PSEUDORANDOM}(\text{seed})$.
     \State Propagate the inflated state in time through
$
    \vb*{y}[j+1] =\vb*{T}\vb*{y}[j] +
      \mqty[\vb*{0} \\\vb*{G } \sqrt{\Delta t} \vb*{w}].
$
\EndFor
\State Assemble the inflated state matrix $
    \vb*{Y} = \left[
\begin{matrix}
  \vert &  &\vert \\
  \vb*{y}[0] & \cdots  &\vb*{y}[N_t]\\
  \vert & &\vert 
  \end{matrix} \right].
$
\end{algorithmic} 
    \begin{algorithmic}[1]
\Function{pseudorandom}{seed}
\State 
The pseudorandom number generator by Burkardt \cite{JBrandom}.
        \EndFunction
        \end{algorithmic}
  \end{algorithm}
        

\begin{algorithm}[ht!]
\caption{Flow-field reconstruction}
  \hspace*{\algorithmicindent} \textbf{Input:}  Inflated-state matrix $\vb*{Y}$, time step $\Delta t$, window function $w(\tau)$ (defaulted as hamming window), SPOD modal basis $\vb*{V}$, model rank $M$, block size $2N_f$. \\
   \hspace*{\algorithmicindent} \textbf{Output:} Reconstructed flow field $\tilde{\vb*{Q}}'$
    \begin{algorithmic}[1]
\For{$i=1:N_t$}
\State Extract the convolutional coordinates $\vb*{a}[i]$ from $\vb*{Y}$.
\State Determine the lower-bound of the reconstruction window as $h_-=1-\min\{i,N_f\}$.
\For{$j=1:2MN_f$}
\State Perform the inversion of convolution in the subspace as 
   \begin{align*}
 \tilde{{a}}_j[i] = \frac{\sum_{h_-}^{0}  \left(a_j[i+h] \right)  \mathrm{e}^{-\mathrm{i}\pi (\lceil j/M\rceil-N_f)  h\Delta t/N_f }}{2N_f\Delta t \sum_{h_-}^{0} w(h\Delta t)}.
   \end{align*}
\EndFor
\EndFor
\State Assemble the convolutional coordinates matrix as 
$$\tilde{\vb*{A}}=\left[
\begin{matrix}
 \tilde{{a}}_1[1] & \cdots & \tilde{{a}}_1[N_t] \\
 \vdots & \ddots &\vdots \\
 \tilde{{a}}_{2MN_f}[1] &\cdots &\tilde{{a}}_{2MN_f}[N_t]
  \end{matrix} \right].$$
\State Reconstruct the spatial-temporal flow field as
$\tilde{\vb*{Q}}'=\vb*{V}\tilde{\vb*{A}}.
  $
    \end{algorithmic}
\end{algorithm}

\section{Convergence and assumption validation} \label{SCGL_validation}

	\begin{figure}[ht]
		\centering
        \includegraphics[trim = 0mm 0mm 0mm 0mm, width=.65\textwidth]{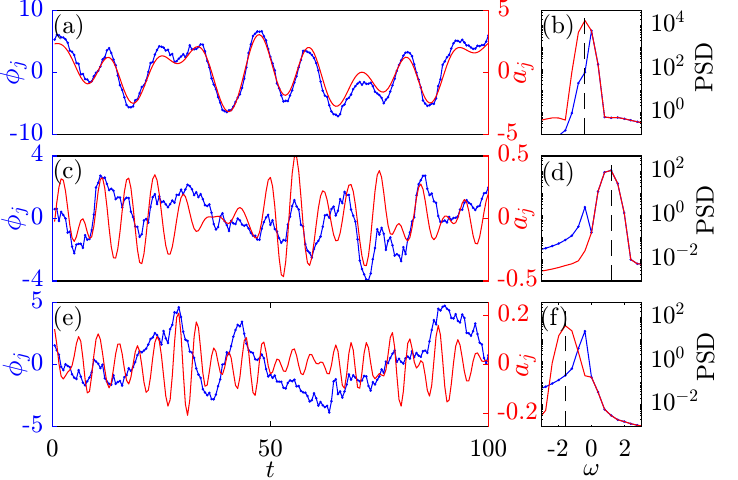}
        \caption{Time traces (a,c,e) and spectra (b,d,f) of the leading SPOD observables $\vb*{\phi}$ (blue) and convolutional coordinates $\vb*{a}$ (red) at three representative frequencies: (a-c) $\omega=-0.39$; (d-f) $\omega=1.18$; (g-i) $\omega=-1.57$. } \label{SCGL_a}
	\end{figure}

We validate the convergence and model assumption of SLICK using the
stochastic complex Ginzburg–Landau equation (SCGL) with spatially-correlated Gaussian white noise as an example.
In this case, arbitrary snapshots are generated via numerical integration, with a baseline total of up to $N_\infty=8\times10^4$ snapshots. 
Figure \ref{SCGL_a} shows the real parts of the leading SPOD observables, $\phi_j=\left<\vb*{\psi}_j,\vb*{q}'\right>_{x}$, and the convolutional coordinates, ${a}_j[i]=  \sum_{h=1-N_f}^{N_f} c_{h,j} \phi_j[i+h] $, at three representative frequencies.
In contrast to $\vb*{\phi}$, the coordinates $\vb*{a}$ are discrete-continuous in time and can effectively capture the intermittent nature of the system states.
 Furthermore, it is noteworthy that the PSD of $\vb*{a}$ peaks at the respective SPOD frequencies, showing that the convolution process preserves the orthogonality property and mode-frequency correspondence inherent in the SPOD.

	\begin{figure}[ht]
		\centering
        \includegraphics[trim = 0mm 0mm 0mm 0mm, width=0.6\textwidth]{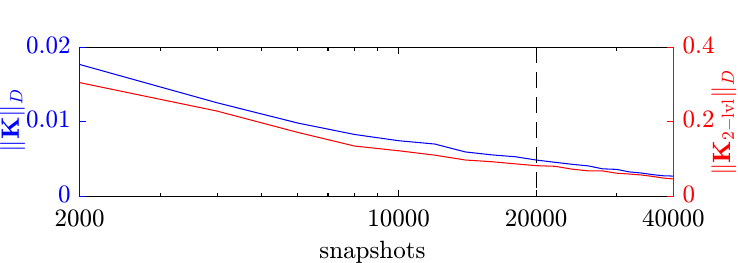}
        \caption{Convergence of the Koopman operators $\vb*{K}$ (blue) and $\vb*{K}_y$ (red).} \label{SCGL_convergence}
	\end{figure}
    
Next, we investigate the convergence of the modified Koopman operator as the amount of data used for its construction increases. 
With a total number of $N_\text{train}$ available training snapshots, we introduce a matrix norm, $\|\cdot \|_D$, as
\begin{subequations}
\begin{align}
    \|\vb*{K}\|_D &\equiv \frac{\|\vb*{K}^{(N_\text{train})}-\vb*{K}^{(N_\infty)}\|_F}{\|\vb*{K}^{(N_\infty)}\|_F},
\end{align}
\end{subequations}
which measures the normalized distance between the matrices constructed with $N_\text{train}$ and $N_\infty$ snapshots.
Figure \ref{SCGL_convergence} shows the convergence of $\vb*{K}=\arg\min_{\vb*{K}} \frac{1}{2} \left( \sum_{i=1}^{N-1} \norm{\dv{}{t}\check{\vb*{a}}[i+1]-\vb*{K}\check{\vb*{a}}[i] }^2 \right)$ and $\vb*{K}_y=  
\left[
\begin{array}{c c}
  \vb*{K} & \vb*{I} \\
 \multicolumn{2}{c}{   \mathrel{\vcenter{\hbox{\rule{0.3cm}{0.6pt}}}}\vb*{M} \mathrel{\vcenter{\hbox{\rule{0.3cm}{0.6pt}}}}}
\end{array}
\right]$, respectively.
As the amount of snapshots increases, both matrices demonstrate nearly algebraic convergence.
The stability of $\vb*{K}_y$ is verified in figure \ref{K_2lvl_eigs}(a).
The training set is then defined as a dataset comprising $N = 20000$ snapshots, where the normalized norm values for the matrices are found to be $\|\vb*{K}\|_D=0.005$ and $\|\vb*{K}_y\|_D=0.082$, respectively. The rest of the snapshots form the test set.



	\begin{figure}[ht]
		\centering
        \includegraphics[trim = 0mm 0mm 0mm 0mm, width=0.5\textwidth]{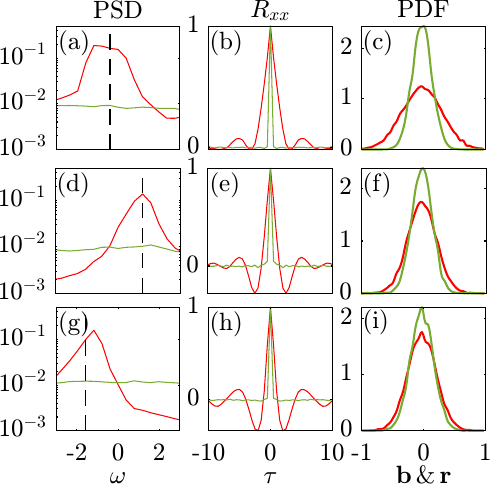}
        \caption{Spectra (left column), autocorrelation estimates (middle column) and the probability distribution (right column) of the rank $2\times 32$ forcing coefficients, $\vb*{b}$ (red), and residue, $\vb*{r}$ (green), at three representative frequencies: (a-c) $\omega=-0.39$; (d-f) $\omega=1.18$; (g-i) $\omega=-1.57$.} \label{SCGL_f_r}
	\end{figure}

Subsequently, we proceed to validate the underlying assumption of the model, which posits that the residue can be modeled as Gaussian random noise.
To accomplish this, we examine the spectra, (normalized) autocorrelations $R_{xx}(\tau) = \left<x(t)x(t+\tau)\right>$, and probability distributions of both the forcing coefficients, $\check{\vb*{b}}(t)$ and the residue, $\check{\vb*{r}}(t)$ in figure \ref{SCGL_f_r}. 
As in panels \ref{SCGL_f_r}(a,d,f), it is evident that the PSD of the residue remains nearly constant across frequencies, whereas the PSD of the forcing coefficients peaks near the corresponding mode frequency.
Panels \ref{SCGL_f_r}(b,e,h) show that the autocorrelations of the residue exhibit a rapid decay, while the forcing coefficients display correlations that persist for 10 or more time units.
 The combined evidence of spectral flatness and the rapid drop of the autocorrelation supports the proposition that the residue can be adequately modeled as white-in-time.
We finally examine the probability distributions of $\check{\vb*{b}}$ and $\check{\vb*{r}}$ in panels \ref{SCGL_f_r}(e,f,i). 
Notably, the probability distributions of residue at different frequencies are nearly Gaussian.
This confirms that $\check{\vb*{r}}$ can be modeled using mutually correlated but white-in-time components.

	\begin{figure}[ht]
		\centering
        \includegraphics[trim = 0mm 0mm 0mm 0mm, width=0.6\textwidth]{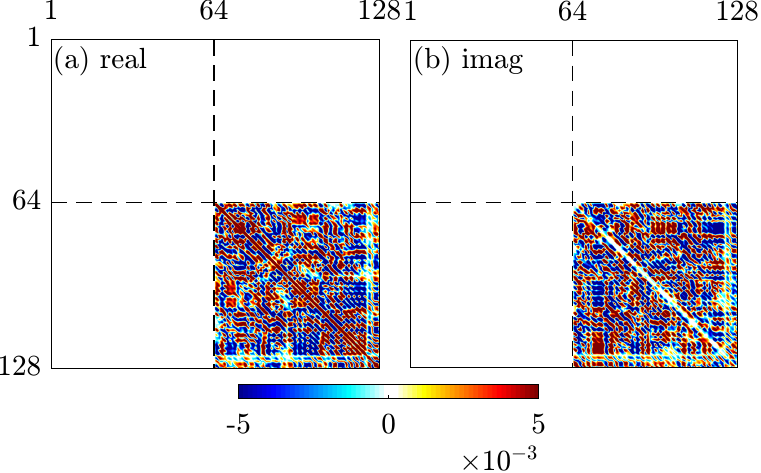}
        \caption{Matrix $\check{\vb*{H}}$: (a) real part; (b) imaginary part.} \label{SCGL_G}
	\end{figure}

    To assess the validity of the dewhitening filter, recall that constructing the matrix $\check{\vb*{H}}=  \check{\vb*{P}}_2 - \vb*{T}   \check{\vb*{P}}_1 \vb*{T}^*$ via the shifted auto-covariance matrices for finite data,
$\check{\vb*{P}}_1 \equiv \mathrm{E}\{ \left(\check{\vb*{y}}\check{\vb*{y}}^* \right)\vert_{i =1}^{N-1} \}$ and $\check{\vb*{P}}_{2} \equiv \mathrm{E}\{ \left(\check{\vb*{y}}\check{\vb*{y}}^* \right)\vert_{i= 2}^{N} \}$, 
is required, as shown in \ref{SCGL_G}.
Notably, both the real and imaginary components of $\check{\vb*{H}}$ are primarily active in the bottom-right quadrant. This observation confirms the viability of truncating $\check{\vb*{H}}$ to solely retain this quadrant.
The matrix $\vb*{G}$ is then obtained as the closest positive semi-definite matrix of $\vb*{H}$ in the Frobenius norm.
This dewhitening filter correlates the white-in-time input ${\vb*{w}}$ for different frequencies, generating the process noise $\widetilde{\vb*{w}}$ that drives the final model.



\section{Eigenvalues of $\vb*{K}_y$}

	\begin{figure*}
		\centering
        \includegraphics[trim = 0mm 0mm 0mm 0mm, clip, width=1\textwidth]{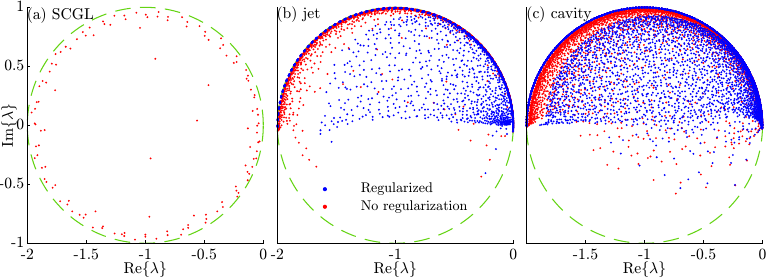}
        \caption{
        Eigenvalues of the inflated Koopman operator, $\vb*{K}_{y}$, for different datasets: (a) SCGL; (b) jet; (c) open cavity flow. The green dashed circle represents the stability region. }\label{K_2lvl_eigs}
	\end{figure*}

 Figure \ref{K_2lvl_eigs} shows the eigenvalues of the inflated Koopman operator $\vb*{K}_y$ for all the cases under consideration. It can be observed that all the eigenvalues stay within the stability region. With L2 regularization, part of the eigenvalues associated with higher frequencies are pushed away from the unit circle. This prevents the overfitting of these high-frequency noise-like signals from the data.



\end{appendix}

\end{document}